\documentclass[preprint,aps,superscriptaddress,nofootinbib,showkeys,11pt]{revtex4-1}
\pdfoutput=1
\usepackage{graphicx}
\usepackage{amsmath,calligra,mathrsfs}
\usepackage{amsfonts}
\usepackage{amssymb}
\usepackage{physics}
\usepackage{tabularx}
\usepackage{multirow}
\usepackage{longtable}
\usepackage{array}
\usepackage{float}
\usepackage{xcolor, soul}
\usepackage{epstopdf}
\usepackage{graphicx}
\usepackage{amsmath}
\usepackage{amsfonts}
\usepackage{amssymb}
\usepackage{xcolor, soul}
\usepackage{epstopdf}
\usepackage{float}
\usepackage{subfigure}
\usepackage{float}
\usepackage{subfigure}
\usepackage{hyperref}
\hypersetup{
     colorlinks   = true,
     citecolor    = red,
     linkcolor    = blue,
     urlcolor     = blue,
}\DeclareGraphicsExtensions{.eps}
\def\beq{\begin{equation}}
\def\eeq{\end{equation}}
\def\bea{\begin{eqnarray}}
\def\eea{\end{eqnarray}}
\def\be{\begin{equation}}
\def\ee{\end{equation}}

\def\bse{\begin{subequations}}
\def\ese{\end{subequations}}

\def\aend{a_{\rm end}}

\def\are{A_{\rm re}}
\def\Are{A_{\rm re}}
\def\aend{a_{\rm end}}
\def\tre{T_{\rm re}}

\def\mp{M_{\rm pl}}
\def\wphi{w_{\rm \phi}}
\def\hend{H_{\rm end}}
\def\wphi{w_{\rm\phi}}
\def\rhor{\rho_{\rm R}}
\def\rhophi{\rho_{\rm \phi}}
\def\tmax{T_{\rm max}}

\def\hf{y_{\rm eff}}
\def\hre{H_{\rm re}}
\def\tfi{T_{\rm fi}}
\def\tfo{T_{\rm fo}}
\def\nss{n_{\rm s}}
\def\ns{n_{\rm s}}
\def\As{A_{\rm s}}
\def\tfio{T_{\rm fi/fo}}
\def\tew{T_{\rm EW}}
\def\sv{\langle \sigma v \rangle}
\def\lmd{\Lambda}
\def\amax{A_{\rm max}}
\def\mx{m_{\rm\chi}}
\def\neq{n_{\rm eq}}
\def\to{T_0}
\def\ar{A_{\rm re}}
\def\mh{m_{\rm h}}
\def\tew{T_{\rm EW}}
\def\omgx{\Omega_{\rm\chi}h^2}
\def\tbbn{T_{\rm BBN}}
\def\pfi{p_{\rm fi}}
\def\po{p_{\rm 0}}
\def\afi{a_{\rm fi}}
\def\ao{a_{\rm 0}}
\def\ath{a_{\rm re}}

\def\phiend{\phi_{\rm end}}
\def\epv{\epsilon}
\def\nk{N_{\star}}
\def\phik{\phi_{\star}}

\usepackage{changes}

\usepackage{cancel}

\graphicspath{{./figs/}}
\begin{document}

\title{Freeze-in and freeze-out production of Higgs portal Majorana fermionic
dark matter during and after reheating}

\author{Rajesh Mondal}%
\email{mrajesh@iitg.ac.in}
\affiliation{%
	Department of Physics, Indian Institute of Technology Guwahati, Assam 781039, India}%
 \author{Sourav Mondal}%
\email{sm206121110@iitg.ac.in}
\affiliation{%
	Department of Physics, Indian Institute of Technology Guwahati, Assam 781039, India}%
\author{Toshifumi Yamada}%
\email{toshifum@mi.meijigakuin.ac.jp}
\affiliation{%
Institute for Mathematical Informatics, Meiji Gakuin University, Yokohama 244-8539, Japan
	}%

\date{\today}

\begin{abstract}
In this paper, we investigate the production of Majorana fermionic dark matter (DM) via the Higgs portal, considering both freeze-in and freeze-out mechanisms during and after the post-inflationary reheating phase. We assume that the Universe is reheated through the decay of the inflaton ($\phi$) into a pair of fermions $f$ and $\bar f$ via the interaction $y\,\phi\,\bar f\,f$, where $y$ is the dimensionless Yukawa coupling. Our analysis focuses on how the non-standard evolution of the Hubble expansion rate and the thermal bath temperature during reheating influence DM production. Additionally, we examine the impact of electroweak symmetry breaking (EWSB), distinguishing between scenarios where DM freeze-in or freeze-out occurs before or after EWSB. We further explore the viable DM parameter space and its compatibility with current and future detection experiments, including XENONnT, LUX-ZEPLIN (LZ), XLZD, and collider searches. Moreover, we incorporate constraints from the Lyman-$\alpha$ bound to ensure consistency with small-scale structure formation.  
\end{abstract}
\maketitle
\section{Introduction}
The Standard Model (SM) of particle physics has successfully explained a wide range of experimental results with remarkable accuracy. However, it fails to address several fundamental mysteries, one of the most significant being the nature of DM. Astrophysical and cosmological observations provide compelling evidence for the existence of non-baryonic DM, which constitutes approximately 27\% of the total energy density of the Universe. For a particle to be a viable DM candidate, it must satisfy several key properties: it should be electromagnetically neutral, colorless, non-relativistic by the time of matter-radiation equality to allow cosmic structure formation, and stable over cosmological timescales. Additionally, its relic abundance must be consistent with the observed value, \(\Omega_{\rm DM} h^2 \simeq 0.12\) \cite{ParticleDataGroup:2020ssz,Planck:2018lbu,Planck:2018vyg}.  

The production of DM in the early Universe can occur through both thermal and non-thermal mechanisms. The most widely studied thermal scenario is the weakly interacting massive particle (WIMP) paradigm \cite{Arcadi:2017kky,Roszkowski:2017nbc}, where DM initially remains in thermal equilibrium with the SM plasma. As the Universe expands and cools, DM decouples from the thermal bath (known as the freeze-out mechanism), giving rise to the observed DM relic abundance by its annihilation cross-section. To achieve the observed DM abundance, the required thermally averaged annihilation cross-section is typically \(\langle \sigma v \rangle \sim 10^{-26} \, \text{cm}^3/\text{s}\) \cite{Steigman:2012nb}. While the WIMP scenario has been extensively explored through direct detection, indirect searches, and collider experiments, the lack of positive signals and increasingly stringent constraints on the WIMP parameter space have motivated the exploration of alternative DM production mechanisms \cite{Arcadi:2017kky}.  

One compelling alternative is the freeze-in mechanism \cite{McDonald:2001vt,Hall:2009bx,Elahi:2014fsa,Bernal:2017kxu}, where DM interacts so feebly with the SM bath that it never reaches thermal equilibrium. Instead, its relic abundance is generated gradually through the decay and annihilations of SM particles. Due to their extremely weak couplings, such DM candidates are often referred to as feebly interacting massive particles (FIMPs). Another simple way to evade the experimental constraints on WIMP is to consider non-standard cosmological histories, such as the post-inflationary reheating epoch. 
In recent years, the study of non-standard cosmological histories has gained substantial interest in DM phenomenology \cite{Maity:2018dgy,Garcia:2020eof,Garcia:2020wiy,Garcia:2021gsy,Giudice:2000ex,Barman:2022tzk,Bhattiprolu:2022sdd,Harigaya:2014waa,Harigaya:2019tzu,Okada:2021uqk,Ghosh:2022fws,Haque:2021mab,Ahmed:2022tfm,Bernal:2022wck,Bernal:2023ura,Bernal:2018kcw,Bernal:2019mhf,Bhatia:2020itt,Hamdan:2017psw,Arcadi:2024wwg,Barman:2024tjt,Gonzalez:2024dtb,Gonzalez:2024rhs,Silva-Malpartida:2024emu,Bernal:2024yhu,Barman:2024mqo,Cosme:2024ndc,Cosme:2023xpa,Haque:2023awl,Haque:2024zdq,Freese:2024ogj,Boddy:2024vgt,Chowdhuryand:2024uvi,Addazi:2017kbx,Chowdhury:2023jft,Banerjee:2024caa}.  


In our work, we investigate the production of Majorana fermionic DM ($\chi$) via the Higgs portal interaction $\bar\chi\chi H^\dagger H/\Lambda$, where \( H \) represents the SM Higgs doublet, and \( \Lambda \) serves as the cutoff scale. While some recent studies \cite{Lebedev:2024vor,Arcadi:2024wwg,Ikemoto:2022qxy,Haba:2024vdg,Biswas:2019iqm} have explored the impact of reheating on Majorana DM production, they often assume instantaneous reheating scenarios and treat the reheating temperature as the maximum bath temperature. However, in realistic inflationary models, reheating occurs over a finite duration, where the maximum temperature can be significantly higher than the reheating temperature \cite{Giudice:2000ex}, and with a distinct thermal evolution during this phase. Since DM production can depend not only on these characteristic temperatures but also on the thermal history of the Universe. Therefore, our analysis adopts a non-instantaneous reheating scenario to capture these effects more accurately. Furthermore, we consider both freeze-in and freeze-out production mechanisms during the reheating phase. During this phase, the inflaton field (\(\phi\)) oscillates around the minima of its potential and gradually decays into SM particles, thereby reheating the Universe. We assume that the inflaton oscillates with a monomial potential, \(V(\phi) \propto \phi^{2\,n}\), leading to an average equation of state (EoS) given by \(\wphi = (n-1)/(n+1)\) \cite{PhysRevD.28.1243}. Depending on the choice of \(n\) and the inflaton decay width, the energy densities of the inflaton and the radiation bath follow a non-trivial evolution during reheating. This modified cosmological history can significantly alter the DM production process, affecting both its relic abundance and detection prospects.  

\textit{This paper is organized as follows}: In Sec.~\ref{sc3}, we provide a detailed discussion of reheating dynamics, establishing the thermal environment for DM production. In Sec.~\ref{DM phenomenology}, we introduce our DM model and examine the impact of reheating on DM production for both freeze-in and freeze-out scenarios. Next, in Sec.~\ref{DM parameter space}, we explore the viable DM parameter space and its compatibility with current and future detection experiments. Finally, we conclude in Sec.~\ref{conclusion}.

\section{\textbf{Reheating dynamics}}\label{sc3}
After the end of inflation, the inflaton field $(\phi)$ starts to oscillate around the minima of its potential, and transfers its energy into massless fields known as radiation. The process of transferring energy from the inflaton field to the radiation is called reheating. Here, we assume that the inflaton potential during reheating takes
the form $V(\phi)\propto\phi^{2\,n}$ with n being an integer, which could originate from the $\alpha$-attractor inflation
models \cite{Kallosh:2013hoa,Kallosh:2013yoa}
\bea
\label{pot1}
V(\phi)=\lambda^4\left(1-e^{-\sqrt{\frac{2}{3\,\alpha}}\frac{\phi}{\mp}}\right)^{2\,n}\,.
\eea
Here, $\mp(=2.4\times10^{18}$ GeV) is the reduced Planck mass, $\lambda$ is the potential scale determined from the CMB measurements \cite{Garcia:2020eof}, and the parameter $(\alpha,n) $ controls the shape of the potential. The coherently oscillating inflaton field can be decomposed as \cite{Shtanov:1994ce,Ichikawa:2008ne}
\[
\phi(t) = \phi_0(t) \mathcal{P}(t),
\]
where \(\phi_0(t)\) represents the time-dependent amplitude of the oscillation, and \(\mathcal{P}(t)\) captures the oscillatory behavior of the inflaton. The oscillatory function can be expressed as a mode expansion,  
\[
\mathcal{P}(t) = \sum_\nu \mathcal{P}_\nu e^{i\,\nu\,\omega\, t},
\]  
where the fundamental frequency is determined to be
\cite{Garcia:2020wiy},
\be \label{fre}
\omega = m_\phi(t)\,\xi\,,~~~\mbox{where}~~\xi=\sqrt{\frac{\pi\,n}{(2\,n-1)}}\frac{\Gamma\left(\frac{1}{2}+\frac{1}{2\,n}\right)}{\Gamma\left(\frac{1}{2\,n}\right)}\,.
\ee
The effective mass of the inflaton is defined as $m_{\phi}^2 ={\partial^2 V(\phi)}/{\partial \phi^2}|_{\phi_0}$ \cite{Garcia:2020eof}, and we have
\begin{eqnarray} \label{mphi}
m_\phi^2\simeq \frac{4n(2n-1)\Lambda^4}{(3\alpha/2)^{n/2}\, M_{\rm pl}^2}\left[\frac{\phi_0}{M_{\rm pl}}\right]^{n-2} 
\sim {(m_\phi^{\rm end})}^2\left[\frac{a}{a_{\rm end}}\right]^{-6\,w_\phi}\,,
\end{eqnarray}
where $\aend$ and $m_\phi^{\rm end}$ denote the scale factor and the inflaton mass at the end of the inflation, respectively. The inflaton mass at the end of inflation is given by
\be\label{massend}
m_\phi^{\rm end}\simeq  \sqrt{\frac{{2n\,\left(2n-1\right)}}{{3\,\alpha}}}\frac{\Lambda^{\frac{2}{n}}}{M_{\rm pl}}\left(\rho_\phi^{\rm end}\right)^{\frac{n-1}{2n}}\,.
\ee
The average energy density of the inflaton ($\rho_{\rm\phi}$) can be defined in terms of the slowly varying amplitude $\phi_0$ as $\rho_{\phi} = (1/2)\langle({\dot \phi}^2 + V(\phi))\rangle = V(\phi)|_{\phi_0}$. During reheating, the inflaton transfers its energy density to SM radiation energy density $(\rho_{\rm R})$. In order to solve reheating dynamics, the Boltzmann equations for the inflaton  and radiation energy density supplemented with the Hubble equation are,
\begin{subequations}{\label{Boltzman}}
\begin{align}
& \dot{\rho_{\rm \phi}}+3H(1+w_\phi)\rho_\phi= -\Gamma_\phi\rho_{\rm \phi}\,(1+w_{\rm \phi}) ,  \label{Boltzman1} \\
&\dot{\rho}_{\rm R}+4H\rho_{\rm R}=\Gamma_{\rm \phi}\rho_{\rm \phi}(1+w_{\rm \phi})\,,
\label{Boltzman2} \\
&H^2=\frac{\rho_{\rm \phi}+\rho_{\rm R}}{3\,M_{\rm pl}^2}\label{Boltzman3}\,,
\end{align}
\end{subequations}
where, dots denote derivatives with respect to cosmic time $t$, and $\Gamma_{\phi}$ is the inflaton decay rate. We consider inflaton decays into a pair of SM fermions $f$ and $\bar f$ through the interaction $y\,\phi\,\bar f\,f$, where $y$ is the dimensionless Yukawa coupling.
Incorporating the inflaton oscillation effect, the effective inflaton decay rate $\Gamma_{\phi}$ can be written as \cite{Garcia:2020wiy,Haque:2023yra,Silva-Malpartida:2023yks,Bernal:2022wck},
\be
\label{Eq:gammaphibis}
\Gamma_{\phi} = 
\frac{y^2_{\rm eff}}{8\,\pi}\,m_{\rm\phi}\,. 
\ee
The effective coupling parameter $y_{\rm eff}$, induced by oscillations, can be calculated as \cite{Garcia:2020wiy,Haque:2023yra,Chakraborty:2023ocr}
\begin{eqnarray}{\label{t2}}
&&y_{\rm eff}^2=(2\,n+2)\,(2\,n-1)\,y^2\,\xi^3\sum^{\infty}_{\nu=1}\nu^3\,\lvert\mathcal P_\nu\rvert^2\,.
\end{eqnarray}
\begin{table}[t]
\caption{Numerical values of the Fourier summations and  the ratio $y_{\rm eff}/y$ for different EoS $\wphi$}\label{fouriersum}
\centering
 \begin{tabular}{||c | c |c |c ||} 
 \hline
 $n(w_\phi)$  & $\sum \nu^3\lvert \mathcal{P}_\nu\rvert^2$  & $y_{\rm eff}/y$\\ [0.5ex] 
 \hline\hline
 1 (0.0) &  $\frac{1}{4}$  & 1\\ 
 2 (1/3)   & 0.241  & 0.71\\
 3 (0.50) & 0.244  & 0.49 \\ [1ex] 
 \hline
 \end{tabular}
\end{table}

The numerical values of $\sum \nu^3\lvert \mathcal{P}_\nu\rvert^2$ for different values of $\wphi$ are provided in table-\ref{fouriersum}. Incorporating all these parameters, we have numerically solved the coupled Boltzmann equation (Eq.~\ref{Boltzman}) with the appropriate initial conditions: \(\rho_\phi(A_{\rm end}=1) = \rho_\phi^{\rm end}=(3/2)\,V(\phi_{\rm end})\,,\rho_R(A_{\rm end}=1) = 0\) (see Appendix-\ref{initcal} for derivation). For the analytical solution, we assume the inflaton decay width remains negligible compared to the Hubble rate throughout the reheating phase. Under this assumption,  the decay term on the right-hand side of Eq.~\ref{Boltzman1} can be neglected, and a straightforward integration yields the following solution,
\begin{eqnarray}\label{rhophi}
&&\rho_\phi(A) \simeq 
\rho^{end}_{\phi}A^{-3(1+w_{\phi})} ,~~~\left[A=a/\aend\right]\,.
\end{eqnarray}
Here, $\rho^{\rm end}_\phi$ is the inflaton energy density at the end of inflation. By using the above solution of Eq.~\ref{rhophi} into Eq.~\ref{Boltzman2}, the solution of radiation energy density $\rho_{\rm R}$ can be written as
\begin{equation}
\rho_{\rm R}(A) \simeq
\frac{\rho_\phi^{\rm end}(1+w_\phi)m^{\rm end}_\phi y_{\rm eff}^2}{4\pi(5-9w_\phi)\,H_{\rm end}}A^{-4}\left[A^{\frac{5-9w_\phi}{2}}-1\right]\,,
\end{equation}
where $H_{\rm end}$ denotes the Hubble constant at the end of the inflation. In thermal equilibrium, the temperature of the inflaton decay products is given by \(\rhor = \epsilon\,T^4\), where \(\epsilon = (\pi^2/30)\,g_\star\), with \(g_\star\) representing the effective number of relativistic degrees of freedom in the thermal bath. For an analytical estimation of the temperature, we assume \(g_\star \sim \mathcal{O}(100)\), which remains a reasonable approximation for temperatures above \(1\) GeV. In our numerical analysis, we solve the following equation along with Eqs.~\ref{Boltzman}, to obtain the correct evolution of the temperature,
\begin{equation}{\label{boltemp}}
    \dot T+\frac{4\,H\,g_\star}{\left(4\,g_\star+T\,\frac{d\,g_\star}{d\,T}\right)}T=\frac{30}{\pi^2}\frac{\Gamma_\phi\,(1+\wphi)}{\left(4\,g_\star+T\,\frac{d\,g_\star}{d\,T}\right)}\frac{\rhophi}{T^3}
\end{equation}
However, for an analytical estimation of the temperature, we take $g_\star$ is constant. At the beginning of the reheating, i.e., at $A=1$, $\rhor=0$ and hence $T=0$. In the instantaneous thermalization
approximation, the temperature of the radiation plasma
initially grows until it reaches a maximum temperature $\tmax$ and then decreases to the temperature at the end
of reheating $\tre$. At $A=A_{\rm max}$ which is typically near to the end of
inflation, $d\rhor/dA=dT/dA=0$, we have 
\begin{eqnarray}\label{amaxnth}
A_{\rm max} &=& \left[\frac{8}{3+9w_\phi}\right]^{\frac{2}{5-9w_\phi}}\,,
\end{eqnarray}

\begin{figure}
    \centering
\includegraphics[width=17.0cm,height=7cm]{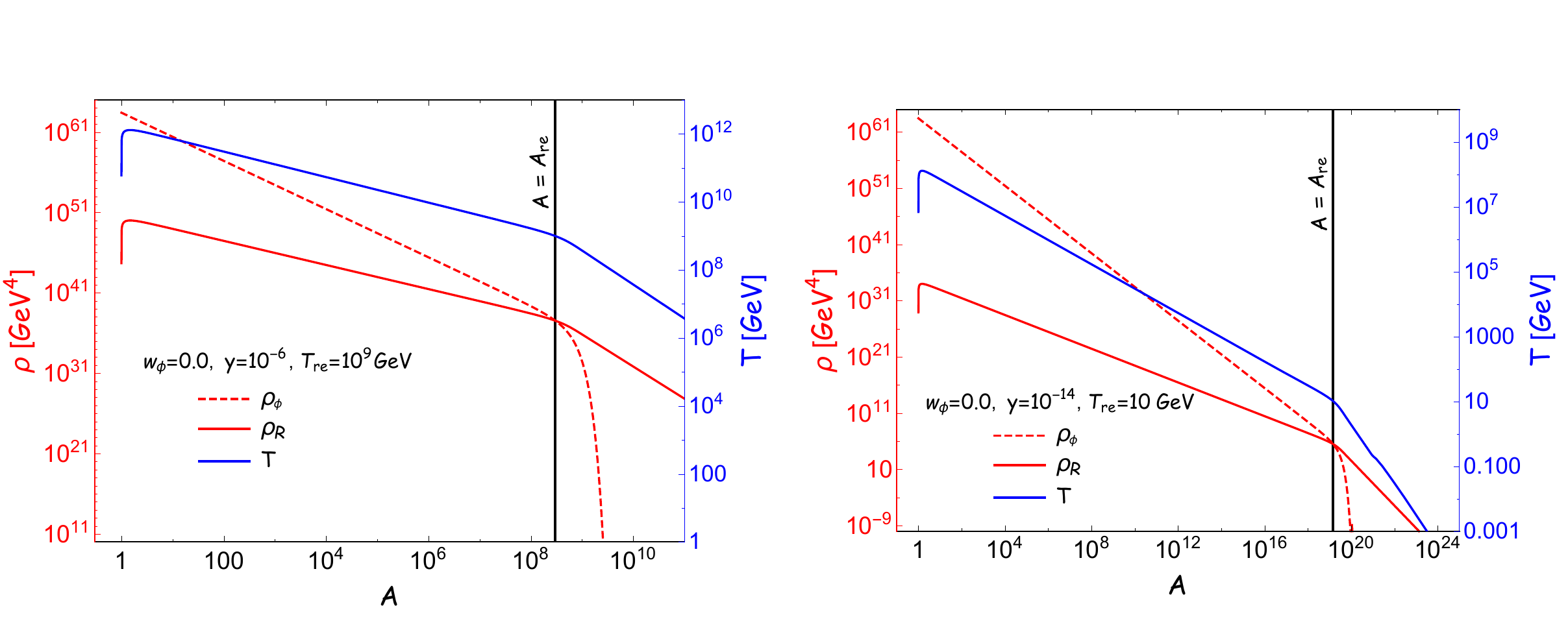}
\includegraphics[width=17.0cm,height=7cm]{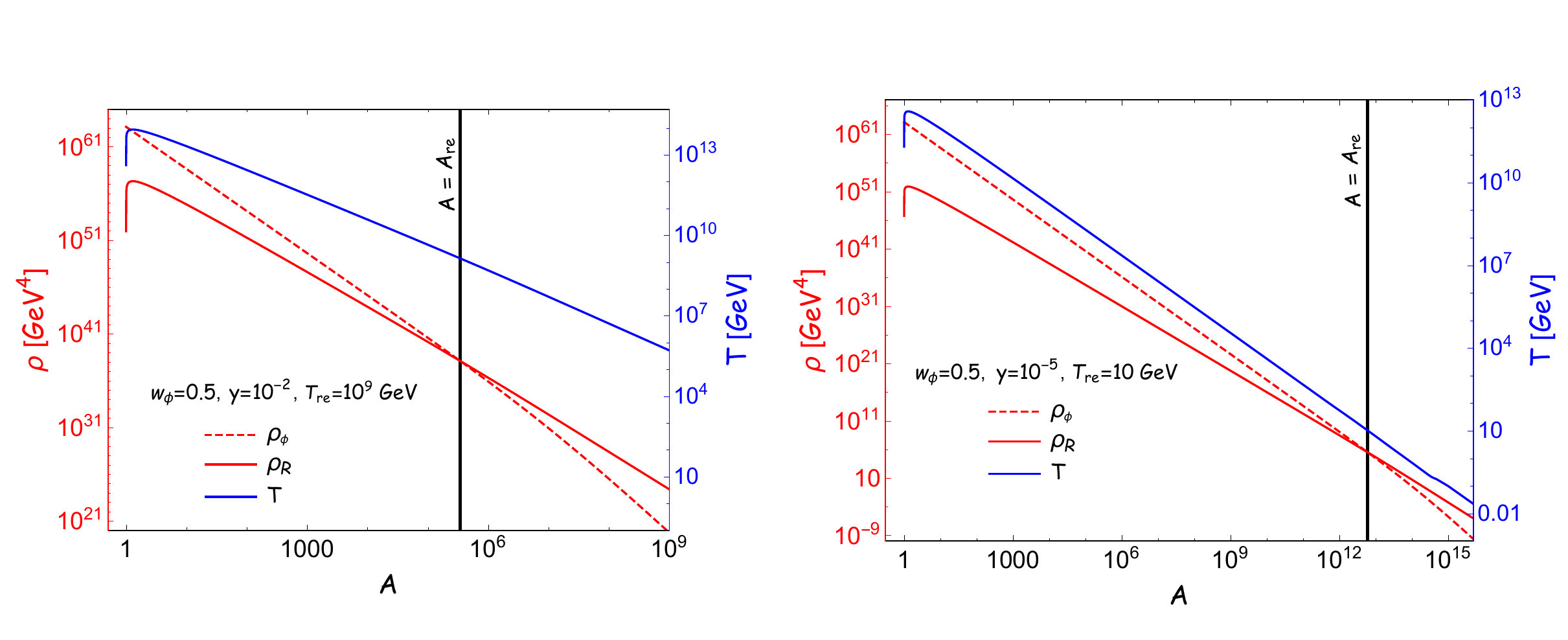}
    \caption{The evolution of the energy densities of the inflaton (dashed red) and radiation (solid red), along with the bath temperature (solid blue), is shown as a function of normalized scale factor $A(=a/\aend)$ for $\tre=10^9$ GeV (left) and $\tre=10$ GeV (right). The top (bottom) panel corresponds to $\wphi=0.0\,(0.5)$, and the vertical black solid lines represent the end of reheating $\are$. }
    \label{inflarad}
\end{figure}
Therefore, the maximum temperature is 
\begin{eqnarray}
\label{mm2}
\tmax &=& 
\left[\frac{6\,\mp^2\,(1+w_\phi)m^{end}_\phi\,H_{end}}{8\,\pi\,\epsilon\,(5-9w_\phi)} y_{\rm eff}^2\right]^{1/4}
\left[\left(\frac{8}{3+9w_\phi}\right)^{\frac{3(1+3w_\phi)}{9w_\phi-5}}-\left(\frac{8}{3+9w_\phi}\right)^{\frac{8}{9w_\phi-5}}\right]^{1/4}
\end{eqnarray}

Now, we will define the most important physical quantities, namely, reheating temperature $T_{\rm re}$, defined at the end of reheating where $\rho_{\phi}(\are) = \rho_R(\are)$,
\begin{align}
T_{re} &= 
\left(\frac{6\mp^2\,(1+w_\phi)m^{end}_\phi H_{end}}{8\pi\epsilon(5-9w_\phi)} \hf^2\right)^{1/4}
\are^{\frac{-3(1+3w_\phi)}{8}},
\end{align}

where
\begin{align}
\are&=
\left(\frac{8\pi(5-9w_\phi)H_{end}}{2(1+w_\phi)m^{end}_\phi\hf^2}\right)^{\frac{2}{3-3w_\phi}}\,
\end{align}
To avoid disrupting Big Bang Nucleosynthesis (BBN), the reheating temperature must satisfy \(\tre\gtrsim\tbbn = 4\) MeV \cite{Kawasaki:2000en,Hannestad:2004px,Barbieri:2025moq,deSalas:2015glj}. The upper bound on the reheating temperature is set by the inflationary scale, with \(\tre \lesssim 10^{15}\) GeV, derived from the constraint on the Hubble parameter during inflation, \(H_{\rm I}^{\rm CMB}\leq 2 \times 10^{-5} \mp\) \cite{Planck:2018jri,BICEP:2021xfz} ({see Appendix-\ref{upperh} for derivation}). However, the thermal bath temperature can be expressed in terms of $\tre$,
 \begin{eqnarray}{\label{trad}}
     T(a)\simeq\tre\left\{\begin{array}{ll}
     &\left(\frac{A}{\are}\right)^{-\frac{3}{8}(1+3\wphi)}~~~\mbox{for}~~A_{\rm max}\leq A\leq \are\,,\\
     &\left(\frac{A}{\are}\right)^{-1}~~~\mbox{for}~~A\geq \are\,.
     \end{array}\right.
     \end{eqnarray}
Similarly, the Hubble parameter can be written as 
     \begin{eqnarray}{\label{hev}}
     H(a)\simeq\hre\left\{\begin{array}{ll}
     &\left(\frac{A}{\are}\right)^{-\frac{3}{2}(1+\wphi)}~~~\mbox{for}~~A_{\rm max}\leq A\leq \are\,,\\
     &\left(\frac{A}{\are}\right)^{-2}~~~\mbox{for}~~A\geq \are\,,
     \end{array}\right.
     \end{eqnarray}
     with Hubble parameter at the end of reheating $\hre\simeq\frac{\pi}{3}\sqrt{\frac{g_\star(\tre)}{10}}\frac{\tre^2}{\mp}$. Note that the expressions of $\tre\,,\are$ and the temperature scaling during reheating (as defined in Eq.~\ref{trad}) are valid only for $\wphi<5/9$. But, when $\wphi>5/9$, radiation production occurs just at the beginning of the reheating, leading $T\propto a^{-1}$ throughout the reheating period; see ref. \cite{Haque:2023yra} for details. 

The evolution of the energy densities of the inflaton and radiation, along with the bath temperature \( T \), as a function of the scale factor, is depicted in Fig.~\ref{inflarad}, for $\tre=(10^9,10)$ GeV. These results are obtained by numerically solving the Boltzmann equations~\ref{Boltzman} together with Eq.~\ref{boltemp}. The top panel corresponds to the case for $\wphi=0$, where the inflaton behaves as matter-like fluid during reheating, following $\rhophi\propto a^{-3}$. Therefore, we recovered the standard scaling of radiation energy density ($\rhor\propto a^{-3/4}$) and temperature $(T\propto a^{-3/8})$ with the scale factor. In the left panel, for temperature evolution, we have found a bump at $T\sim0.15$ GeV due to the QCD phase transition. The bottom panels correspond to \(\wphi=0.5\), where the inflaton oscillates in a sextic potential, causing its energy density to redshift as \(\rhophi \propto a^{-9/2}\). Consequently, the radiation energy density evolves as \(\rhor \propto a^{-15/4}\), and the temperature scales as \(T \propto a^{-15/16}\). Notably, this temperature evolution closely resembles the standard cosmological scaling \(T \propto a^{-1}\), leading to a smooth transition from reheating to the radiation-dominated era without any significant change in slope.
\section{Dark matter phenomenology}\label{DM phenomenology}
 In this section, we study the implications of reheating dynamics on the production of DM. In our analysis, we consider both freeze-in and freeze-out mechanisms for DM production. For DM production, we introduce a gauge-singlet Majorana fermion, $\chi$, to the SM field content. Additionally, we impose $Z_2$ symmetry, under which $\chi$ is odd, and the SM fields are even. The relevant part of the Lagrangian involving $\chi$ can be written as \cite{Ikemoto:2022qxy} :
\begin{equation}
    \mathcal{L}\supset \frac{1}{2}\bar\chi\,(i\gamma^\mu\,\partial_\mu-\mx)\chi-\frac{1}{\Lambda}\bar\chi\chi H^\dagger H\,,
    \label{DMlagrangian}
\end{equation}
where \( H \) represents the SM Higgs field, and \( \Lambda \) serves as the cutoff scale. Due to the imposed \( Z_2 \) symmetry, the neutral particle \( \chi \), with a mass \( m \), is stabilized, making it a natural candidate for DM.

\begin{figure}[h] 
 	\begin{center}
 		\includegraphics[width=13.0cm,height=3.0cm]{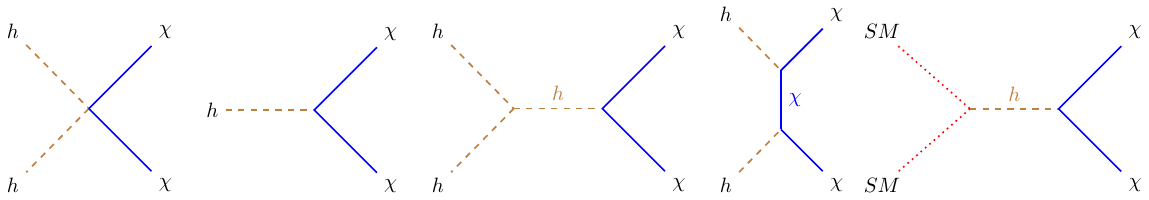}\quad
 	\end{center}
\caption{Feynman diagrams illustrating the dominant DM production channels.}\label{feynmand}
 \end{figure}

Before EWSB, the Higgs field is in a symmetric phase with no vacuum expectation value (VEV), which means the masses of the SM particles are zero. In this high-temperature regime, the Higgs boson itself is massless, and interaction is totally governed by the unbroken $SU(2)_L \times U(1)_Y$ gauge symmetry. So before EWSB, DM particles are produced only through Higgs annihilation via contact diagram $HH^\dagger\rightarrow\chi\chi$. However, after EWSB i.e. $T<T_{\rm EW}$, the Higgs boson acquires VEV($v$) and expanding the contact operator $\bar\chi\chi H^\dagger H$ around $v$ as $H= v + h$, we have
\begin{equation}
    \frac{1}{2\,\Lambda}(h^2+2\,v\,h)\bar\chi\,\chi+...\,.
\end{equation}
The first term represents the usual four-point interaction, and the last term generates the $3$-point vertex $h\rightarrow\chi\chi$, which not only accounts for the direct Higgs decay but also enables massive SM particles to annihilate into DM through $s$-channel Higgs exchange, $\chi$- mediated $t$- and $u$- channel (see Fig.\ref{feynmand}). Depending on the freeze-in/out temperature (\(\tfio\)), there are two possible scenarios:  
(A) DM production occurs before EWSB (\(\tfio > \tew\)) and  
(B) DM production occurs after EWSB (\(\tfio < \tew\)). We will first discuss Case A, followed by Case B.

\subsection{DM production before EWSB}
If the DM freeze-in/out temperature ($T_{\rm fi/fo}$) is higher than the electroweak temperature ($\tew$), the DM production is completed before the EWSB. Before EWSB,  DM particles are produced only through Higgs annihilation via contact diagram $HH\rightarrow \bar\chi\chi$. The annihilation cross-section for this process can be written as
\begin{equation}
    \sigma(s)= \frac{1}{32\,\pi\,s\, \Lambda^2}\frac{\left(s-4\,m_{\chi}^2\right)^{\frac{3}{2}}}{\left(s-4\,m_h^2\right)^{\frac{1}{2}}}
\end{equation}
To track the evolution of the DM, we solve the following Boltzmann equation
\begin{equation}{\label{dmeq}}
\frac{dn_{\chi}}{dt} + 3Hn_{\chi} = \langle \sigma v \rangle \left(n_{\text{eq}}^2 - n_{\chi}^2 \right),
\end{equation}
where $n_{\rm eq}$ is the equilibrium number density, and $\sv$ is the thermally averaged cross section times velocity, which is defined as \cite{Gondolo:1990dk} :
\begin{equation}
    \langle\sigma v\rangle=\frac{g_{\rm\chi}^2}{n^2_{\rm eq}}\frac{T}{32\pi^4}\int^{\infty}_{4m^2_{\rm \chi}}\sqrt{s}(s-4m^2_{\rm \chi})\sigma(s)K_1\left(\frac{\sqrt{s}}{T}\right)ds\,,
\end{equation}
 where $g_{\rm\chi}=2$ is the internal
degrees of freedom for the Majorana fermion. Using the above two equations, one can find the following expression for $\sv$
\begin{equation}{\label{sigmavv}}
    \sv = \frac{g_{\rm\chi}^2}{n_{\text{eq}}^2} \cdot \frac{T^6}{32\,\pi^5\,\Lambda^2}\, I_1(m_{\chi}, T)
\end{equation}
where,
\begin{equation*}
    I_1(\mx, T) = 
    \begin{cases} 
        \frac{3\pi}{8} \frac{m_{\rm{\chi}}^2}{T^2} e^{-\frac{2 m_{\rm{\chi}}}{T}} \left( 1 + \frac{7}{4} \frac{T}{m_{\rm{\chi}}} \right) & \text{for } \mx > T, \\
        1 - \frac{3}{4} \frac{m_\chi^2}{T^2} & \text{for } \mx<T.
    \end{cases}
\end{equation*}


Depending on the strength of the coupling parameter \(\lmd\), the production of DM particles \(\chi\) can occur via either the freeze-in or freeze-out mechanism. In the following sub-section, we discuss these two different production mechanisms in detail.
\subsubsection{Freeze-in}
\begin{figure}
    \centering
\includegraphics[width=0.85\linewidth]{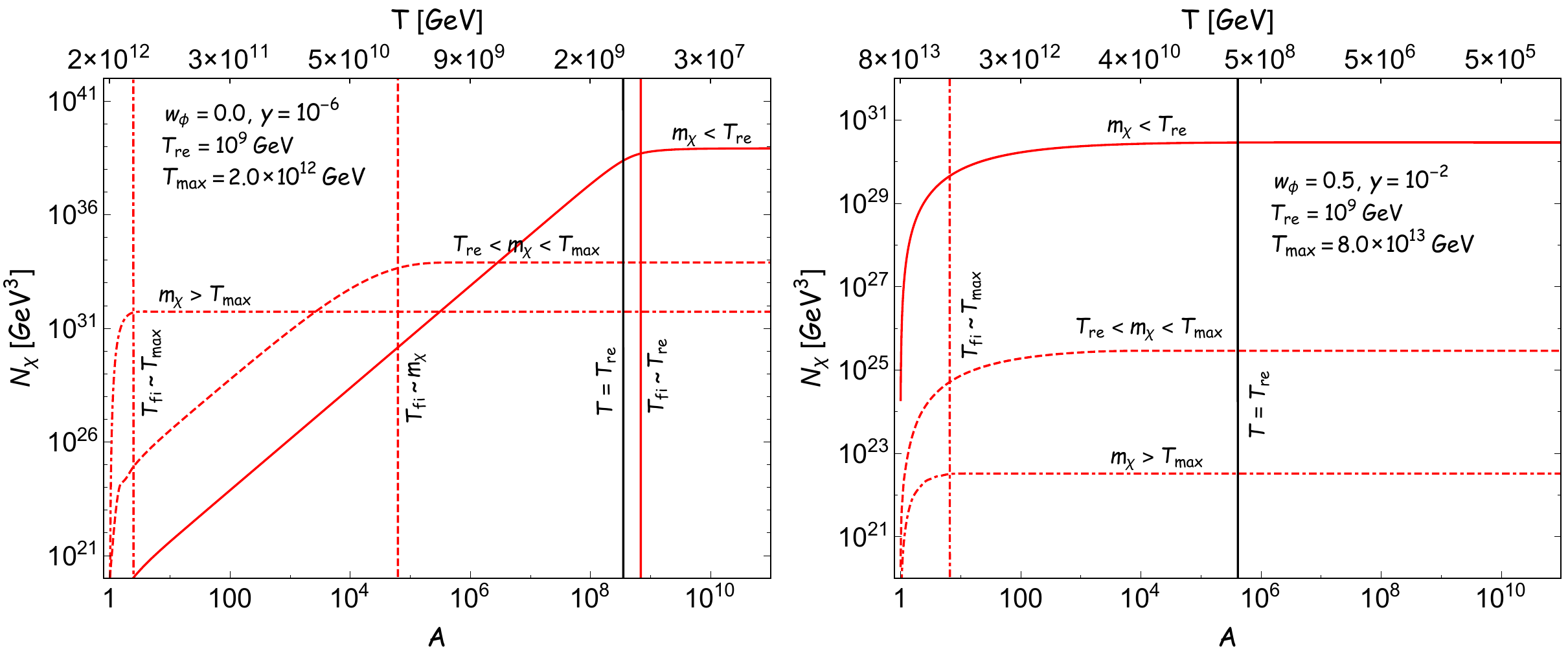}
    \caption{The evolution of the co-moving number density of DM for freeze-in as a function of the scale factor $A$ (bottom x-axis) and the bath temperature $T$ (top x-axis) for $\wphi=0.0$ (left), $\wphi=0.5$ (right) with $\tre=10^{9}$ GeV. Each plot presents three different cases, \textit{Left panel}: $\mx (=10^6\,\mbox{GeV})<\tre$ (solid line, $\Lambda=8.4\times10^{18}$ GeV), $\tre<\mx (=10^{11}\,\mbox{GeV})<\tmax$ (dashed line, $\Lambda=3.6\times10^{16}\,\mbox{GeV}$) and $\mx(=5.0\times10^{12}\,\mbox{GeV})>\tmax$ (dot-dashed, $\Lambda=2.0\times10^{12}\,\mbox{GeV}$); \textit{Right panel}: $\mx (=10^6\,\mbox{GeV})<\tre$ (solid line, $\Lambda=7.2\times10^{19}$ GeV), $\tre<\mx (=10^{11}\,\mbox{GeV})<\tmax$ (dashed line, $\Lambda=2.0\times10^{22}\,\mbox{GeV}$) and $\mx(=1.5\times10^{15}\,\mbox{GeV})>\tmax$ (dot-dashed, $\Lambda=7.5\times10^{16}\,\mbox{GeV}$). The vertical red line corresponds to $T\sim \tfi$, and the vertical black line corresponds to $T=\tre$. }
    \label{ficoming}
\end{figure}

For freeze in production, the DM can never reach thermal equilibrium, i.e. $n_\chi\ll n_{\rm eq}$; hence the co-moving number density $N_\chi = n_\chi A^3$ follows the following simple equation,
\begin{equation}{\label{relbeq}}
    \frac{dN_\chi(A)}{dA}=\frac{A^2}{H}\langle\sigma v\rangle (n_{\rm eq})^2 .
\end{equation}
Usually, for freeze-in cases, the DM production continues to happen until the radiation temperature equals the DM mass $T\simeq m_\chi$. So, if the DM mass is greater than $\tre$, i.e. $m_\chi>\tre$, the freeze-in happens during the course of reheating. Therefore, in order to solve the above equation, DM can be safely assumed to be relativistic, and the equilibrium number density becomes, 
\begin{equation}{\label{relneq}}
    n_{\rm eq}=\frac{g_{\rm\chi}\,T^3}{\pi^2}
\end{equation}
Using the Eqs. \ref{trad} ,\ref{hev},~\ref{sigmavv} and \ref{relneq} in Eq. \ref{relbeq}, one can obtain the following solution of the DM,
\begin{equation}{\label{bmgt}}
   N_{\rm{\chi}}(A) =\frac{g_{\rm\chi}^2}{24\,(3-7\,\wphi)\,\pi^5 \,\Lambda^2}\frac{A^3_{\text{re}}\tre^6}{\hre} \left[\left( \frac{A}{A_{\text{re}}} \right)^{\frac{3(3-7w_\phi)}{4}} -\left( \frac{\amax}{A_{\text{re}}} \right)^{\frac{3(3-7w_\phi)}{4}} \right]
\end{equation}
From the above solution, $3-7\wphi>0$ when $\wphi<3/7$, therefore most of the production occurs around $A\sim A\,(T\sim \mx)$, consequently, the freeze-in temperature $\tfi\sim m_\chi$\footnote{This type of freeze-in scenario is 
known as infra-red (IR) freeze-in \cite{Bernal:2017kxu,Yaguna:2011qn,Chu:2011be,Blennow:2013jba,Merle:2015oja,Shakya:2015xnx,Hessler:2016kwm,Konig:2016dzg,Duch:2017khv,Heeba:2018wtf,Lebedev:2019ton}, where bulk of the DM production occurs at $T\sim\mx$.}. But, on the other hand, $3-7\wphi<0$ when $\wphi>3/7$, most of the production occurs around $A\sim\amax$ and this leads to the freeze-in temperature $\tfi\sim T_{\rm max}$ \footnote{ DM production that occurs predominantly at very high temperatures, typically near the maximum temperature of the Universe, is referred to as Ultraviolet (UV) freeze-in \cite{Hall:2009bx,Elahi:2014fsa, 
  McDonald:2015ljz, 
  Chen:2017kvz, 
  Biswas:2019iqm, 
  Bernal:2019mhf, 
  Bernal:2020bfj, 
  Bernal:2020qyu, 
  Barman:2020plp,Barman:2021tgt}.}. These two scenarios are illustrated in Fig. \ref{ficoming} (dashed lines). In the left plot, for $\wphi=0$, the co-moving number density is frozen at $m_\chi/T\sim 1$, and in the right side, for $\wphi=0.5$, the co-moving number density is frozen at the beginning of reheating (at $T\sim\tmax$).

Since $\tre$ is not the maximum temperature, the above solution (Eq. \ref{bmgt}) is valid within $\tmax>\mx>\tre$, but not valid for $\mx>\tmax$. For the $\mx>\tmax$ limit, DM production has Boltzmann suppression from the beginning of reheating. Using $\sv$ and $\neq$ for $\mx>T$ limit, one can obtain the following solution
\begin{equation}
     N_{\rm{\chi}}(A) =\frac{g_{\rm\chi}^2\,\mx}{64\,(1+3\,\wphi)\,\pi^4 \,\Lambda^2}\frac{A^3_{\text{re}}T^5_{\text{re}}}{H_{\text{re}}} \left(\frac{T_{\text{re}}}{T_{\text{max}}}\right)^{\frac{7-11\wphi}{1+3\wphi}} e^{-\frac{2\,\mx}{\tmax}}
\end{equation}
For $\mx > \tmax$, most of the production occurs near $A \sim \amax$, therefore the freeze-in temperature $\tfi \sim T_{\rm max}$. This behavior is clearly evident from Fig.~\ref{ficoming} (dot-dashed lines).

So far, we have discussed the DM production when DM mass is larger than $\tre$. Furthermore, if we consider the DM mass to be less than the reheating temperature (i.e. $\mx<\tre$ ), expecting that the freeze-in happens after the reheating is over. For this case, the solution of co-moving number density can be written as :
\begin{equation}
     N_{\rm{\chi}}(A) =\frac{g_{\rm\chi}^2}{32\,\pi^5 \,\Lambda^2}\frac{A^3_{\text{re}}T^6_{\text{re}}}{H_{\text{re}}} \left( 1-\frac{A_{\text{re}}}{A}\right)
\end{equation}

Although $\mx<\tre$, the DM production continues up to $T\sim\tre$ point, then its production becomes negligible. This is true for $\wphi<3/7$ (see solid line in the left Fig.~\ref{ficoming}) and the freeze-in temperature $\tfi\sim\tre$. On the other hand, when $\wphi>3/7$, the DM production is frozen just at the beginning of reheating, which we discussed earlier (see solid line in the right Fig.~\ref{ficoming}) and the freeze-in temperature $\tfi\sim\tmax$ .
\begin{figure}
    \centering
\includegraphics[width=17.0cm,height=5cm]{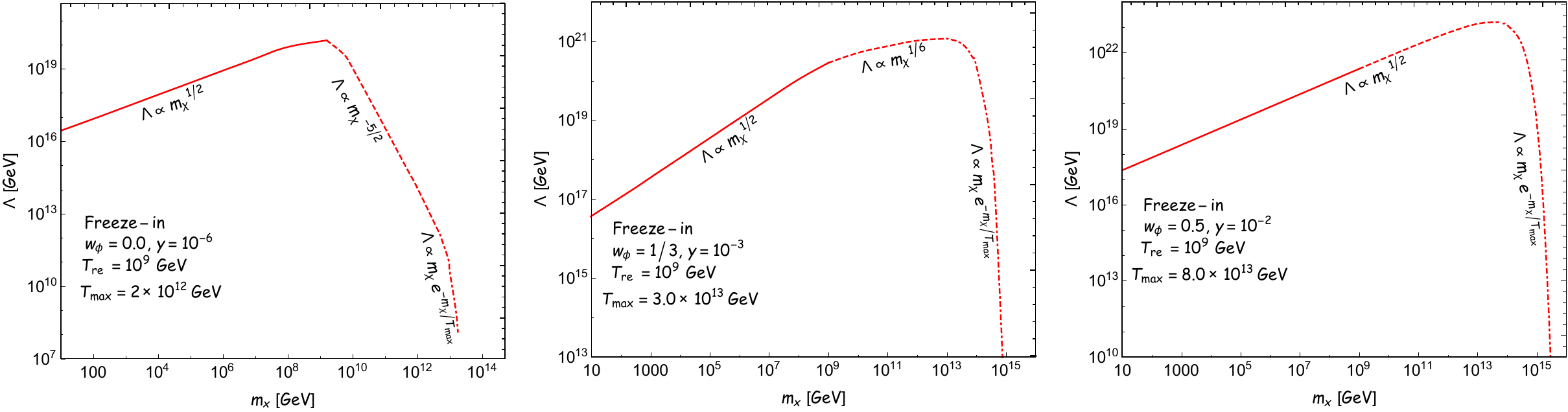}
    \caption{The behavior of the DM parameter space \((\Lambda, \mx)\) for the freeze-in mechanism in the pre-EWSB scenario, consistent
with the observed relic abundance. Here, we consider three different EoS $\wphi=0.0$ (left), $1/3$ (middle), and $0.5$ (right), with a fixed $\tre=10^9$ GeV. Different segments of the line correspond to distinct mass regimes: solid ($\mx<\tre$), dashed ($\tre<\mx<\tmax$), and dot-dashed ($\mx>\tmax$). The region above the line corresponds to an underabundant DM scenario, where the predicted relic abundance is below the observed value.}
    \label{casebfi}
\end{figure}
After the freeze-in, the co-moving DM number density is constant, which gives the correct DM relic abundance. The expression of the present-day relic abundance for all the cases are
\begin{eqnarray}\label{wx6}
\Omega_{\rm{\chi}}=\Omega_{R} h^2 \frac{\sqrt{3}\mp\,g_{\rm\chi}^2}{24\,\pi^5\epsilon^{\frac{3}{2}}\,\to} \left\{\begin{array}{ll}
&\frac{m_{\chi} T_{\text{re}}}{(3-7w_{\phi})\, \Lambda^2}\left( \frac{\mx}{\tre} \right)^{\frac{ 14\,w_{\phi}-6}{1 + 3w_{\phi}}}\,,~~~\mbox{for}~~~\tmax> \mx>\tre\,,\wphi<3/7\\
&\frac{m_{\chi} T_{\text{re}}}{(3-7w_{\phi}) \,\Lambda^2}\,,~~~\mbox{for}~~~ \mx<\tre\,,\wphi<3/7\\
& \frac{m_{\chi} T_{\text{re}}}{(3-7w_{\phi})\, \Lambda^2} \left( \frac{\tmax}{\tre} \right)^{\frac{(7\,w_{\phi}-3)}{4}}\,,~~~\mbox{for}~~~ \mx<\tmax\,,\wphi>3/7\\
& \frac{3\pi}{8}\frac{m_{\chi}^2} {(1+3w_{\phi}) \Lambda^2}\left(\frac{T_{\text{re}}}{T_{\text{max}}}\right )^{\frac{7-11\wphi}{1+3\wphi}}e^{-\frac{2\,\mx}{\tmax}} 
 \,, ~~~\mbox{for}~~\mx>\tmax\,.
\end{array}\right.
\end{eqnarray}

In Fig.~\ref{casebfi}, we have shown the behavior of the DM parameter space ($\Lambda\,,\mx$) for freeze-in, consistent with the observed relic abundance $\Omega^{\rm obs}_\chi h^2\simeq0.12$. This figure corresponds to the pre-EWSB scenario, considering three different EoS $\wphi=0.0$ (left), $1/3$ (middle), and $0.5$ (right), with a fixed $\tre=10^9$ GeV. Depending on the DM mass scale, we identify three distinct regimes: (a) $\mx<\tre$: $\Omega_xh^2\propto\mx/\Lambda^2$ for a fixed $\tre$, which leads to the slope $\Lambda\propto\mx^{1/2}$. This behavior is recovered in Fig.~\ref{casebfi} (see solid line). (b) $\tre<\mx<\tmax$: in this case $\Omega_xh^2$ scales as $\Lambda^{-2}\,\mx^{(17\wphi-5)/(1+3\wphi)}$ when $\wphi<3/7$, which accounts for the slope $\Lambda\propto\mx^{-2.5} \,(\mx^{1/6})$ for $\wphi=0.0\,(1/3)$ in the Fig.~\ref{casebfi} (dashed line). As a result, we see that the required value of $\Lambda$
decreases (increases) with increasing (decreasing) DM mass $\mx$ for $\wphi=0.0\,(1/3)$. Again, when $\wphi>3/7$, the abundance $\omgx\propto\mx\,\Lambda^2$ same as $\mx<\tre$, as a result, no change in the slope is observed for $\wphi=0.5$ (c) $\mx>\tmax$, in this regime, the relic abundance follows the scaling $\Omega_xh^2\propto\mx^2\Lambda^{-2}e^{-2\mx/Tmax}$, implies that $\Lambda\propto\mx\,e^{-\mx/Tmax}$. The required value of $\Lambda$ decreases very rapidly as $\mx$ increases (see the dot-dashed line in Fig.~\ref{casebfi}).

\subsubsection{Freeze-out}
Depending on the strength of the interaction cross-section, DM can be produced through either thermal or non-thermal processes. In case of thermal production, which is known as the freeze-out mechanism, the cross-section has to be much larger compared to the freeze-in mechanism in order to reach the thermal equilibrium. The freeze-out temperature $\tfo$ is defined as :
\begin{equation}{\label{fo1}}
    \langle\sigma v\rangle\, n_{\rm eq}(\tfo)=H(\tfo) .
\end{equation}
During reheating, the inflaton decays into SM particles, leading to non-conservation of entropy. This physical situation gives rise to  two distinct scenarios:

(a)~{\underline{Freeze out after reheating ($\tfo<\tre$)}}: This is known as the standard freeze-out scenario, where DM freeze-out occurs well after reheating, in the standard radiation-dominated era. After the freeze-out, the comoving number density $N_\chi=n_\chi A^3$ will be much larger than the comoving equilibrium number density $N_{\rm eq}^r$. Therefore, once freeze-out occurs, $N_{\rm eq}$
  can be neglected\footnote{Our main results are based on a complete numerical solution of the
Boltzmann equation Eq.~\ref{dmeq}, while the analytical solution of Eq.~\ref{relbeq1} is provided to give
intuitive, physical insight into the scaling behavior of DM abundance.} in comparison to $N_{\rm {\chi}}$ and from the Eq.~\ref{dmeq},
\begin{equation}{\label{relbeq1}}
    \frac{dN_\chi(A)}{dA}=-\frac{\langle\sigma v\rangle}{H}A^{-4}\,N^2_\chi .
\end{equation}
 Taking into account the radiation-dominated era and integrating from the freeze-out point ($\tfo$) to the present time ($T_{0}$), we get the present-day comoving number density ($N_{0}$)
 \begin{equation}
     N_0=\frac{16\,\pi\,\Lambda^2}{3}\,\frac{m_\chi\,\hre}{\tre}\Are^3\left(\frac{\tre}{\tfo}\right)^2=\frac{16\,\pi\,\sqrt{\epsilon}\,\Lambda^2}{3\,\sqrt{3}}\,\frac{\mx\,\tfo}{\mp}\,A^3_{\rm fo}\,.
 \end{equation}
We replaced $\hre$ in terms of $\tre$ (using Eq.~\ref{hev}), to obtain the final expression. The freeze-out temperature $\tfo$ can be written as (from Eq. \ref{fo1}),
 \begin{equation}\label{tfopost}
     \tfo=2\,\mx\,\frac{1}{W_{-1}\left[2\,\mx\,\mathcal{K}^{-2}_0\right]}~~~\mbox{where}~~~\mathcal{K}_0=\frac{16\,\pi^2\,\Lambda^2}{3\,g_\chi\,\mp}\left(\frac{2\,\epsilon\,\pi}{3\,\mx}\right)^{1/2}\,,
 \end{equation}
 where $W_{-1}$ is the Lambert function with branch $-1$. Therefore, the current abundance can be written as,
 \begin{equation}
     \Omega_\chi h^2=\Omega_{\rm R} h^2\frac{\mx}{\epsilon\,\to}\frac{N_0}{\tre^3\,\ar^3}=\Omega_{\rm R} h^2\frac{16\,\pi}{3\,\sqrt{3\,\epsilon}}\frac{\Lambda^2}{T_0\,\mp}\left(\frac{m_\chi}{\tfo}\right)^2\,.
 \end{equation}
Interestingly, both the freeze-out temperature and relic abundance are intriguingly independent of the reheating temperature $\tre$, a remarkable contrast to the freeze-in mechanism where $\tre$ plays a pivotal role. This confirms that the standard freeze-out scenario is independent of the dynamics of reheating.\\
\begin{figure}
    \centering
\includegraphics[width=0.85\linewidth]{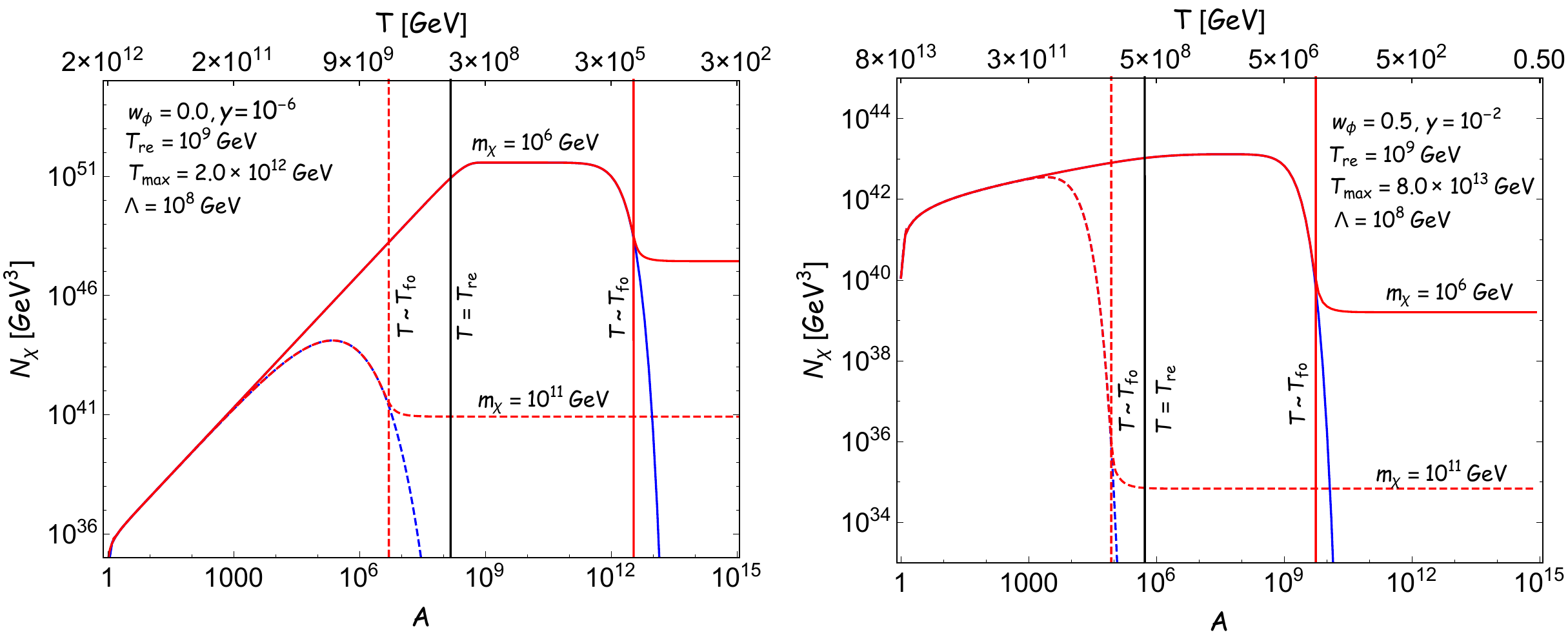}
    \caption{The evolution of the comoving DM number density for freeze-out as a function of the scale factor $A$ (bottom x-axis) and the bath temperature $T$ (top x-axis) for $\wphi=0.0$ (left), $\wphi=0.5$ (right) with $\tre=10^{9}$ GeV. Each plot presents two different cases: freeze-out after reheating $\tfo<\tre$ (solid line), and freeze-out during reheating $\tfo>\tre$ (dashed line). In both plots, $\Lambda=10^8$ GeV. The vertical red line corresponds to $T\sim \tfo$, and the vertical black line corresponds to $T=\tre$. }
    \label{focoming}
\end{figure}

(b)~{\underline{Freeze out during reheating ($\tfo>\tre$)}}:
 While dark matter freeze-out is typically assumed to occur after reheating, during the radiation-dominated era, it can also take place during reheating. In this scenario, the freeze-out dynamics would depend on the non-trivial temperature evolution during reheating, adding complexity to the thermal history of the Universe. By solving the Eq. \ref{relbeq1}, during reheating from the freeze-out point ($\tfo$) to the end of reheating ($\tre$), the co-moving number density at the end of reheating is given by :
 \begin{equation}
     N_{\rm{\chi}}(\tre)\simeq\frac{4\,\pi\,(15-3\,\wphi)\,\Lambda^2}{12}\frac{\mx\,\hre}{\tre}\left(\frac{\tfo}{\tre}\right)^\frac{5-\wphi}{1+3\,\wphi}\,.
 \end{equation}
Therefore, the present-day DM relic abundance can be expressed as :
\begin{equation}{\label{abundance}}
    \Omega_\chi h^2=\Omega_{\rm R} h^2\frac{\mx}{\epsilon\,\to}\frac{N(\tre)}{\tre^3\,\ar^3}\simeq\Omega_{\rm R} h^2\frac{\pi\,(5-\wphi)}{\sqrt{3\,\epsilon}}\frac{\Lambda^2}{\to\,\mp}\left(\frac{\mx}{\tre}\right)^{2}\left(\frac{\tfo}{\tre}\right)^\frac{\wphi-5}{1+3\,\wphi}\,,
\end{equation}
and the freeze-out temperature $\tfo$ is 
\begin{equation}
    \tfo=\frac{2\,\mx}{5-2\,j}\frac{1}{W_{-1}\left[\frac{2\,\mx}{5-2\,j}\,\mathcal{K}^{\frac{2}{2\,j-5}}\right]}~~~~\mbox{where}~~~\mathcal{K}=\frac{8\,\pi(2\,\pi)^{3/2}}{3\,g_{\rm\chi}}\frac{\Lambda^2\,\hre}{\sqrt{\mx}}\tre^{-j}\,,j=\frac{4\,(1+\wphi)}{1+3\,\wphi}\,.
\end{equation}
The evolution of the co-moving DM density for freeze-out is shown in Fig.~\ref{focoming}, as a function of the scale factor $A$ (bottom x-axis) and the bath temperature $T$ (top x-axis). We have taken two different values of $\wphi=0.0$ (left plot) and $\wphi=0.5$ (right plot) with $\tre=10^{9}$ GeV.
\begin{figure}
    \centering
\includegraphics[width=17.0cm,height=5cm]{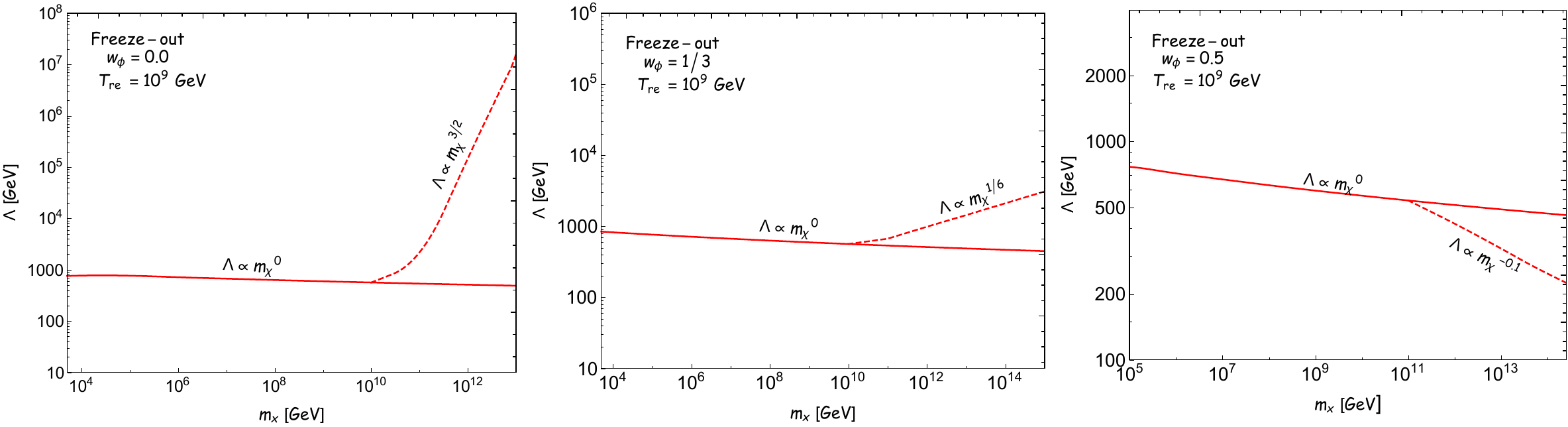}
    \caption{The behavior of the DM parameter space \((\Lambda, \mx)\) for the freeze-out mechanism in the pre-EWSB scenario, consistent
with the observed relic abundance. Here, we consider three different EoS $\wphi=0.0$ (left), $1/3$ (middle), and $0.5$ (right), with a fixed $\tre=10^9$ GeV.  The solid red line corresponds to the standard freeze-out scenario, where DM freezes out in the radiation-dominated epoch, and the impact of reheating is shown with red dashed line. The region below the line corresponds to an underabundant DM scenario, where the predicted relic abundance is below the observed value. }
    \label{casebfo}
\end{figure}

To show the behavior of DM parameter space for freeze-out in the pre-EWSB scenario, we have plotted the coupling scale \(\Lambda\) as a function of DM mass $\mx$ in Fig.~\ref{casebfo}, ensuring it reproduces the observed relic abundance. We classify the freeze-out scenarios based on whether it occurs before or after reheating:\\
(a) Freeze-out after reheating (\(\tfo < \tre\)): In this case, dark matter freeze-out takes place well after reheating, provided that the DM mass satisfies \(\mx < 25\,\tre\) \cite{Bernal:2022wck,Silva-Malpartida:2023yks}. This condition follows from the fact that freeze-out typically occurs at a temperature $T_{\rm fo} \simeq m_\chi / x_f$, with $x_f \sim 20$-25. The resulting relic abundance, \(\Omega_\chi h^2\), is independent of the inflaton equation of state (\(\wphi\)), reheating temperature (\(\tre\)), and DM mass (\(\mx\)). Although $m_\chi$ appears explicitly in the expression of $T_{\rm fo}$ (Eq.~\ref{tfopost}), the Lambert function ensures that $T_{\rm fo}$ scales approximately linearly with $m_\chi$. Consequently, in expressions such as Eq.~(33), where $(m_\chi / T_{\rm fo})^2$ appears, the explicit dependence on $m_\chi$ effectively cancels. This explains why the relic abundance $\Omega_\chi h^2$ is approximately independent of the DM mass \footnote{with a small logarithmic dependence on the DM mass}, despite the apparent appearance of $m_\chi$ in the formula. The observed relic abundance is typically achieved for a coupling scale of \(\Lambda \simeq \mathcal{O}(10^3)\) GeV, irrespective of the DM mass (solid line).\\
(b) Freeze-out during reheating (\(\tfo > \tre\)): If the DM mass is \(\mx\gtrsim25\,\tre\) \cite{Bernal:2022wck,Silva-Malpartida:2023yks}, freeze-out occurs during the reheating phase. In this case, the relic abundance scales as \(\Omega_\chi h^2 \propto \Lambda^2\, \mx^{(7\wphi - 3)/(1 + 3\wphi)}\) (from Eq.~\ref{abundance}), assuming \(\tfo \propto \mx\). Consequently, the slope of the \(\Lambda\) vs. \(\mx\) curve varies depending on the inflaton EoS \(\wphi\); for \(\wphi = 0.0\), the coupling scale follows \(\Lambda \propto \mx^{3/2}\), meaning that \(\Lambda\) increases with \(\mx\). For \(\wphi = 1/3\), the relation becomes \(\Lambda \propto \mx^{1/6}\), showing a milder increase. However, for \(\wphi = 0.5\), the trend reverses, with \(\Lambda\) decreasing as \(\mx\) increases, following \(\Lambda \propto \mx^{-0.1}\). All these behaviors are illustrated with red dashed lines in Fig.~\ref{casebfo}.

We note that some of the dark matter (DM) masses shown in Fig.~\ref{casebfo}, presented for illustrative purposes, exceed the unitarity bound for thermal WIMPs, $m_\chi\gtrsim 100~\mathrm{TeV}$. These examples are included solely to illustrate the behavior of the DM parameter space during freeze-out after reheating. Nevertheless, the qualitative features of the $\Lambda$-$m_\chi$ relation remain valid even beyond the unitarity limit.

\subsection{DM production after EWSB}
If the DM freeze-in/out temperature ($T_{\rm fi/fo}$) is smaller than the electroweak temperature ($\tew$), DM production from the thermal bath continues even after EWSB. After EWSB, the Higgs doublet acquires a VEV, which enables additional channels for DM production, as discussed earlier. So, the Boltzmann equations associated with
DM takes the following form
\begin{equation} {\label{darkd1}}
\dot n_{\chi}+3Hn_{\chi}+\langle\sigma v\rangle( n_{\chi}^2- n_{\rm eq}^2)+\langle\Gamma_{h}\rangle n_{\rm h} \left(1-\frac{n^2_{\chi}}{n^2_{\rm eq}}\right)=0\,,
\end{equation}
where $n_{\rm h}$ is the equilibrium Higgs number density, and here, $\langle\sigma v\rangle$ is the total thermally averaged annihilation cross-section, accounting for all the relevant processes (see Fig.~\ref{feynmand}). 
$\langle\Gamma_{h}\rangle$ is the thermally average decay width $h\rightarrow \chi\chi$  which is zero before EWSB , an it can be written as :
\begin{equation}
\langle\Gamma_{h}\rangle=\Gamma_{\rm h\rightarrow \chi\chi}\frac{K_{1}\left(\frac{m_{h}}{T}\right)}{K_{2}\left(\frac{m_{h}}{T}\right)},~~~\mbox{where}~~~\Gamma_{\rm h\rightarrow \chi\chi}=\frac{m_{\rm h}\,v^2}{4\,\pi\,\Lambda^2}\left(1-\frac{4\,\mx^2}{m^2_{h}}\right)^{3/2}
\end{equation}
where $K_{n}$ is the modified Bessel function of the $n$-th order and $m_{h}=125$ GeV is the Higgs mass. 
\begin{figure}
    \centering
\includegraphics[width=17.0cm,height=6cm]{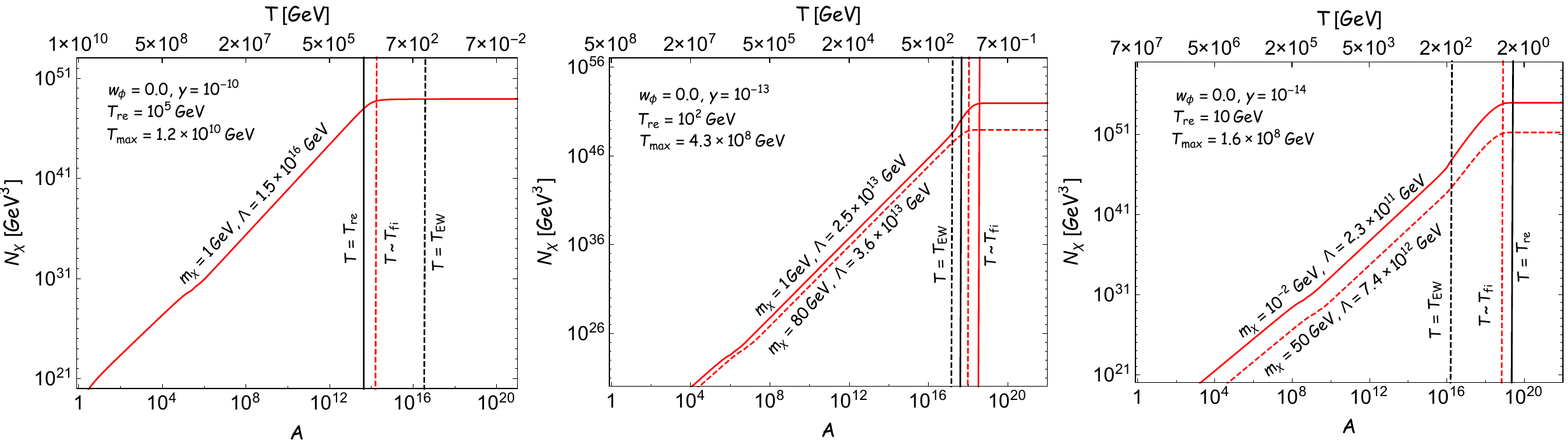}
    \caption{The evolution of the co-moving DM number density for freeze-in mechanism as a function of the scale factor $A$ (bottom x-axis) and the bath temperature $T$ (top x-axis) for $\wphi=0.0$. The vertical red lines indicate \(T \sim \tfi\), while the vertical solid and dashed black lines represent \(T = \tre\) and \(T = T_{\rm EW}\), respectively.}
    \label{comovingcase}
\end{figure}
\subsubsection{Freeze-in}
The DM production depends on the inflaton EoS \(\wphi\), leading to two different possibilities. When \(\wphi > 3/7\), DM production predominantly occurs at the very onset of reheating, with negligible contributions from later stages, including those after EWSB. Therefore, EWSB has no impact on freeze-in DM production if $\wphi>3/7$. The evolution of the DM number density and the present-day relic abundance follow the same expressions as derived for the pre-EWSB case.

On the other hand, when $\wphi<3/7$,the impact of EWSB  depends on the reheating temperature, leading to  three different possibilities:
\begin{itemize}
    \item $\bf{\tre>T_{\rm EW}}$ : DM production is completed within the reheating phase if \(\mx > \tre\). When \(\mx < \tre\), one might expect production to continue until \(T \sim \mx\); however, in reality, it ceases at \(T \sim \tre\). Consequently, for $\mx<\mh/2$, even though one might anticipate DM production to persist after EWSB, it actually stops at $T=\tre$. For instance, as shown in the left Fig.~\ref{comovingcase}, we illustrate the evolution of the comoving number density for a DM mass of \(\mx = 1\) GeV, assuming \(\tre = 10^{5}\) GeV. Although, DM can be produced after EWSB via both 2-to-2 scatterings and the Higgs decay channel \(h \rightarrow \chi\chi\), its production is effectively completed by the end of reheating, at \(T \sim \tre\). Therefore, for the case where $\tre>T_{\rm EW}$, DM freeze-in always occurs before EWSB for all DM masses, and thus, EWSB does not play any role. Consequently, its relic abundance follows the same expression as derived earlier in Eq.~\ref{wx6}.
    \item $\bf{m_{\rm h}/2<\tre<T_{\rm EW}}$ : In this case, DM production exhibits qualitatively different behavior depending on whether DM mass $\mx$ is  above and below $\mh/2$. When $\mx>\mh/2$,  DM is solely produced by 2-to-2 scatterings, while for $\mx<\mh/2$, dominant contribution coming from  Higgs decays ($h\rightarrow\chi\chi$). However, unlike the previous case, if the  $\mx<\tre$, the DM production continues after reheating up to $T\sim\mx\,(T\sim\mh/2)$ for DM mass heavier (lighter)  than $\mh/2$. Consequently, in this scenario, we find that EWSB plays a crucial role when the DM mass is below $\mathcal{O}(500)$ GeV. In Fig.~\ref{comovingcase} (middle panel), we illustrate the evolution of the comoving number density for DM masses $\mx=80$ GeV (dashed line) and $\mx=1$ GeV (solid line), assuming \(\tre = 10^{2}\) GeV. Since both chosen DM masses are below the threshold limit ($\mathcal{O}(500)$ GeV), their production naturally continues after EWSB, even after reheating as $\mx<\tre$. Therefore, the freeze-in occurs when the respective conditions-\(T \sim \mx\) for \(\mx > m_h/2\) and \(T \sim m_h/2\) for \(\mx < m_h/2\) are satisfied.
    \item $\bf\tre<m_{\rm h}/2$ : In this scenario, EWSB  also plays a crucial role, and the corresponding DM mass to be below $\mathcal{O}(500)$ GeV. For this case, freeze-in always occurs during reheating, regardless of the DM mass $\mx>\tre$ or $\mx<\tre$. In Fig.~\ref{comovingcase} (right panel), we show the evolution of the comoving number density for DM masses $\mx=50$ GeV (dashed line) and $\mx=10^{-2}$ GeV (solid line), assuming \(\tre = 10\) GeV. In both cases, the DM production is completed at $T\sim\mh/2$ within reheating.
\end{itemize}
\begin{figure}[h] 
 	\begin{center}
\includegraphics[width=14.0cm,height=11cm]{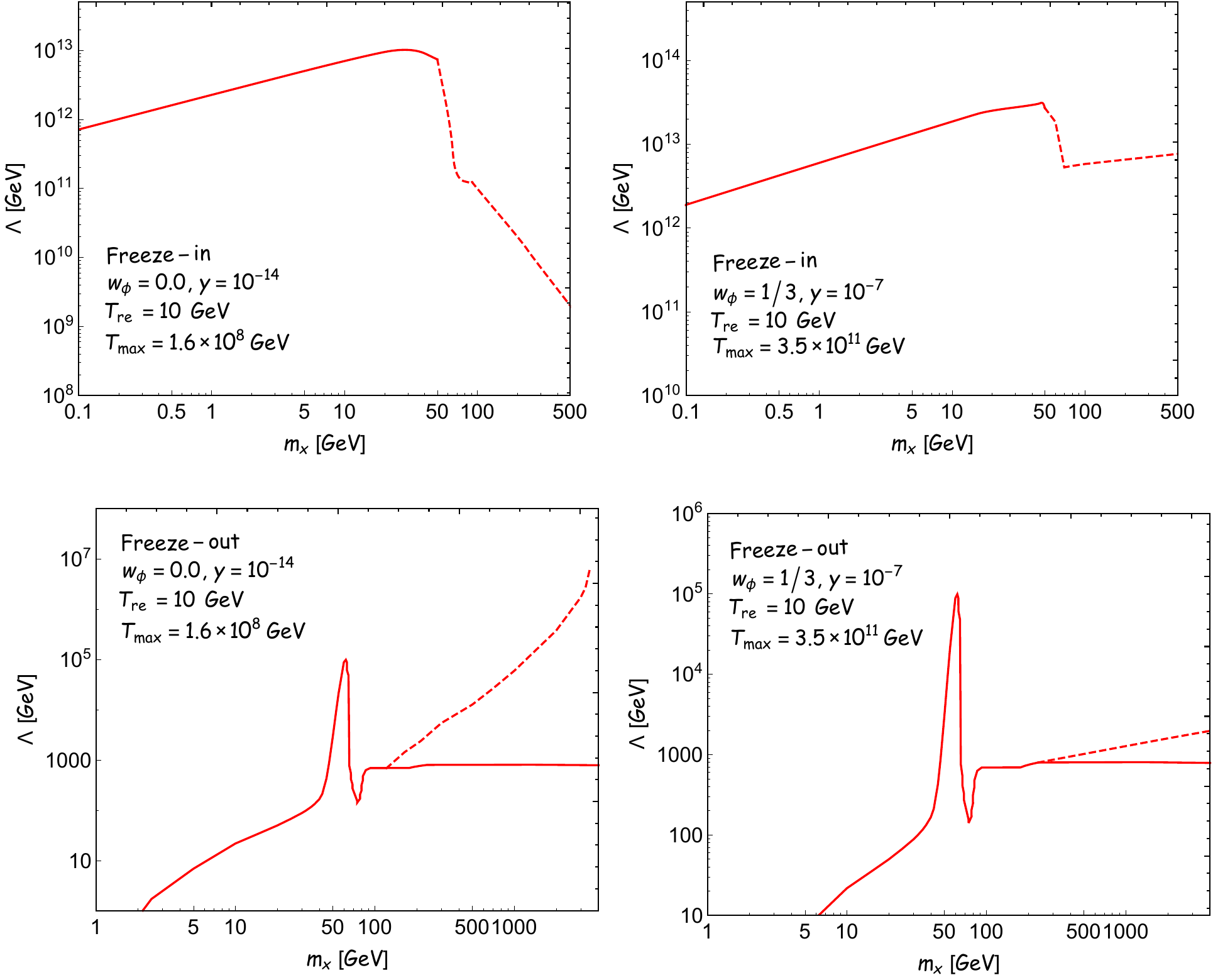}\quad
 \end{center}
  \caption{The behavior of the DM parameter space \((\Lambda, \mx)\) for the post-EWSB scenario, consistent
with the observed relic abundance. The top (bottom) panel corresponds to the freeze-in (freeze-out) for $\wphi=0.0$ (left) and $\wphi=1/3$ (right). For freeze-in, the solid (dashed) line corresponds to the DM mass $\mx$ heavier (lighter) than $\mh/2$. For freeze-out, the solid red line corresponds to the standard
freeze-out scenario and while  the red dashed line illustrates the impact of reheating.}
\label{casecombo}
  \end{figure}
We have analyzed the behavior of the DM parameter space \((\Lambda, \mx)\) in Fig.~\ref{casecombo} (top panel) for the freeze-in mechanism in the post-EWSB scenario. We found that if \(\wphi < 3/7\) and reheating temperature \(\tre < \tew\), EWSB can affect freeze-in production only if the DM mass below \(\mathcal{O}(500)\) GeV. In Fig.~\ref{casecombo}, we consider $\tre=10$ GeV ($<\tew$) for two different EoS $\wphi=(0,1/3)$. A striking qualitative difference emerges between DM masses above and below \(\mh/2\), with the most noticeable feature being the sharp drop at the Higgs resonance \(\mx \approx \mh/2\). For \(\mx < \mh/2\), DM is dominantly produced via Higgs decays, leading to an expected relic abundance proportional to \(\mx\), as freeze-in occurs around \(T \sim \mh/2\) and the slope is $\Lambda\propto\mx^{1/2}$ (see solid line). In contrast, for \(\mx > \mh/2\), production is dominated by 2-to-2 scatterings, which changes the slope of the \(\Lambda\) vs. \(\mx\) relation (dashed line). Due to the complexity of the interactions, an analytical estimation of the slope in this regime is not possible. Note that when $\wphi>3/7$, all DM masses are produced before EWSB irrespective of $\tre$, and its behavior is the same as the pre-EWSB case, therefore we do not show the parameter space for $\wphi=0.5$.
\subsubsection{Freeze-out}
We have illustrated the behavior of the DM parameter space \((\Lambda, \mx)\) in Fig.~\ref{casecombo} for the freeze-out mechanism in the post-EWSB scenario. This parameter space is obtained by numerically solving Eq.~\ref{darkd1} together with Eq.~\ref{Boltzman}. We find that any DM mass $\mx\gtrsim\mathcal{O} (4)$ TeV freezes out before EWSB; otherwise, freeze-out occurs after EWSB. Thus, the EWSB scenario can significantly influence the DM freeze-out, if the DM mass is below $\mathcal{O} (4)$ TeV, for any $\wphi$ and $\tre$. The impact of reheating on freeze-out production is shown with red dashed lines. The solid red line corresponds to the standard freeze-out scenario, in which  DM freezes out in the radiation-dominated epoch.
\begin{figure}[h] 
 	\begin{center}
\includegraphics[width=13.50cm,height=15.5cm]{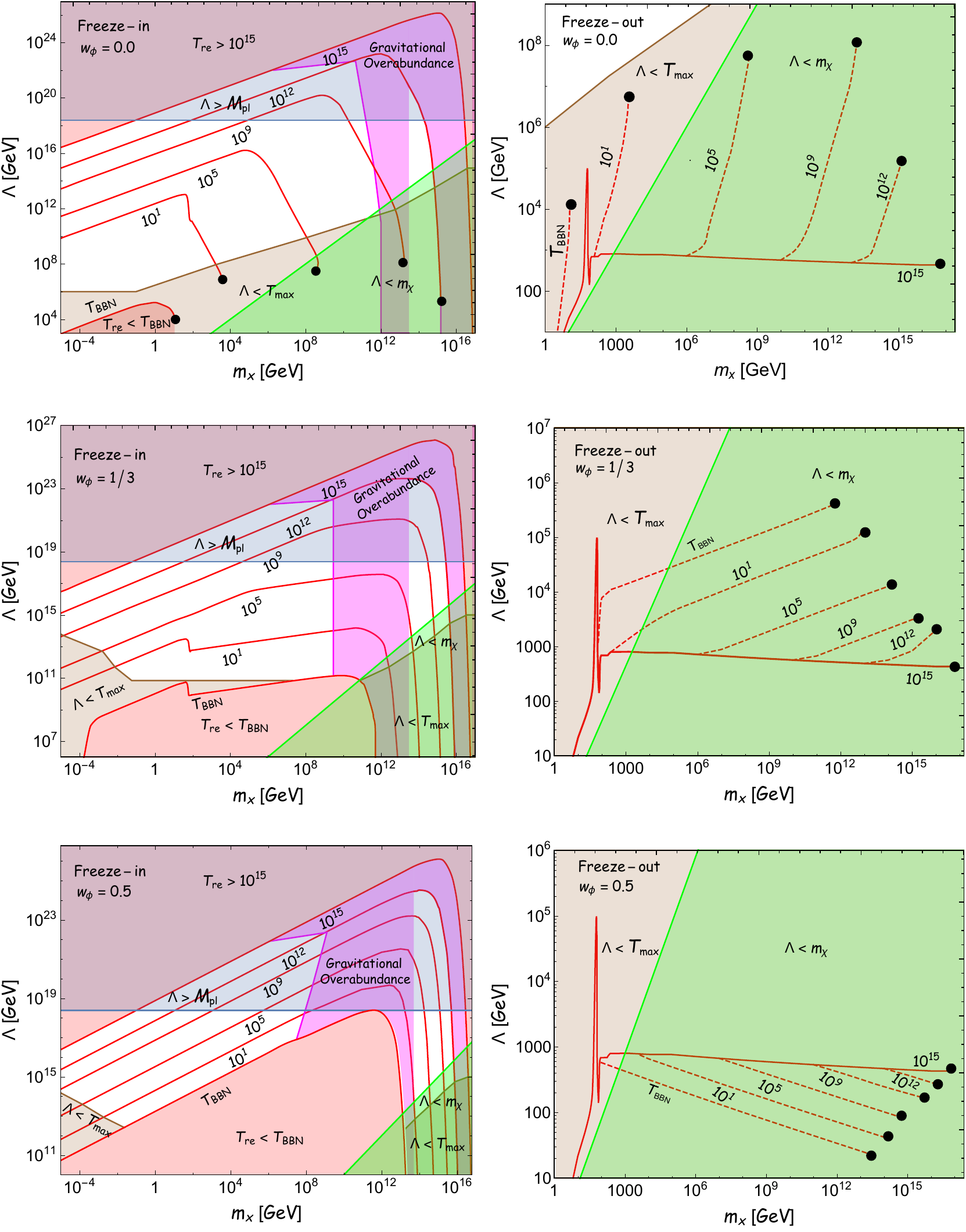}\quad
 \end{center}
  \caption{The complete DM parameter space \((\Lambda, \mx)\) is shown for both freeze-in (left panel) and freeze-out (right panel) scenarios, with different values of the EoS: \(\wphi=0.0\) (top), \(\wphi=1/3\) (middle), and \(\wphi=0.5\) (bottom). Along the red lines, the correct DM relic abundance is achieved for a fixed reheating temperature \(\tre\). The small black circle represents
the freeze-in and freeze-out coincidence point. The colored shaded regions correspond to various constraints, as discussed in the text. }
\label{fullpara}
  \end{figure}
\section{DM parameter space and its constraints}\label{DM parameter space}
In the previous section, we provided a detailed analysis of DM production and the behavior of its parameter space. In this section, we extend our discussion to explore the complete DM parameter space along with its relevant constraints. Figure.~\ref{fullpara} illustrates the complete DM parameter space, where we consider six distinct reheating temperatures: \( \tre = (\tbbn, 10, 10^5, 10^9, 10^{12}, 10^{15}) \) GeV, covering the entire range from the minimum to the maximum reheating temperatures. The colored shaded regions represent different constraints. The red-shaded region is excluded by the constraints on the reheating temperature \( \tre \). The light blue region represents the parameter space where the Higgs-Majorana interaction is Planck-suppressed (\(\Lambda > \mp\)), supported by the WIMPZILLA hypothesis \cite{Kolb:2017jvz}. Since $\tmax$ is the maximum temperature during reheating, the effective theory of DM described in Eq.~\ref{DMlagrangian} is valid only if \( \Lambda \geq \tmax \) \cite{Garcia:2020eof,Barman:2024tjt}; the brown-shaded region violates this condition. This constraint rules out the entire freeze-out parameter space (see right plot), while in the freeze-in scenario, only certain regions of the DM parameter space are excluded. Furthermore, the unitarity constraint $\Lambda > m_\chi$ imposes an additional constraint, excluding the green-shaded region. The magenta-shaded region is discarded \footnote{For the freeze-out case, the DM parameter space is also affected by universal gravitational production, especially at very high masses \cite{Haque:2023yra}. However, since the entire freeze-out scenario is ruled out by the condition $\Lambda > \tmax$, we have not included it.} due to the gravitational overproduction of DM \cite{Clery:2021bwz,Haque:2022kez}.For gravitational production, we have considered contributions from both inflaton and thermal bath gravitational annihilations. The gravitational production from the thermal bath can significantly contribute to the present DM relic abundance if $\tre \geq \mathcal{O}(10^{12})$ GeV \cite{Clery:2021bwz}. Additionally, DM masses below \( \mathcal{O}(10^{-5}) \) GeV are excluded by Lyman-\(\alpha\) \cite{Irsic:2017ixq,Ballesteros:2020adh} constraints for freeze-in (see Appendix-\ref{Lymanbound}).The small dots represent the points where the freeze-in and freeze-out solutions meet, i.e., where there is a smooth transition from freeze-in to freeze-out or vice versa. Below this point, DM attains chemical equilibrium and is produced thermally via the freeze-out mechanism, whereas above it, chemical equilibrium is never established, and DM is generated non-thermally via the freeze-in mechanism. The corresponding mass at this point is the maximum DM mass capable of producing the correct relic abundance. For masses above this value, the DM becomes underabundant today.

\begin{figure}
    \centering
\includegraphics[width=17.0cm,height=4.80cm]{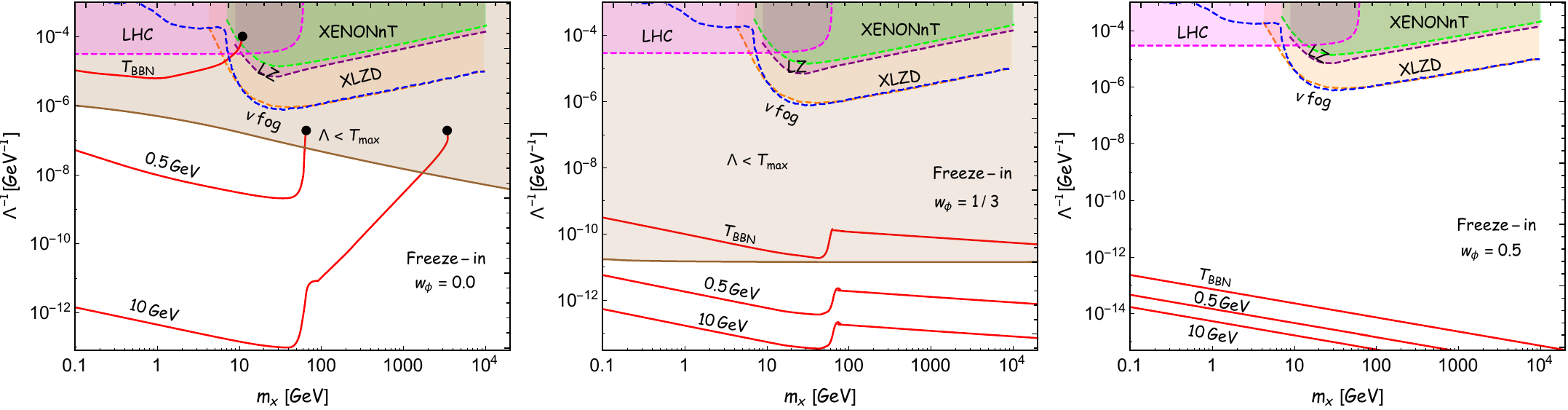}
    \caption{Comparison of the freeze-in DM parameter space \((\Lambda^{-1}, \mx)\) with the sensitivity of current and future detection experiments. The three red lines correspond to different reheating temperatures: \(\tre = (\tbbn, 0.5, 10)\) GeV. Shaded regions indicate experimental sensitivities: XENONnT (green), LZ (purple), XLZD (orange), and LHC constraints (magenta), while the brown shaded area represents excluded parameter space ($\lmd<\tmax$). The blue dashed line (``$\nu$ fog'') marks the neutrino background limit for direct DM detection."}
    \label{detection}
\end{figure}

In Fig.~\ref{detection}, we compare our DM parameter space \((\Lambda^{-1}, \mx)\) with the sensitivity of current and future DM search facilities. Specifically, we include constraints from XENONnT \cite{XENON:2023cxc}, LUX-ZEPLIN (LZ) \cite{LZ:2022lsv}, and the projected reach of XLZD \cite{Aalbers:2022dzr}. The effective spin-independent cross-section \(\sigma_{\rm SI}\) for elastic DM-nucleon scattering is given by  
\begin{equation}{\label{directdet}}
    \sigma_{\rm SI}=\frac{4\,\mu_{\rm N}^2\,f_{\rm N}^2\,m_{\rm N}^2}{\pi\,\lmd^2\,m_h^4}\,,
\end{equation}  
where $\mu_{\rm N}=m_{\rm N}\,\mx/(m_{\rm N}+\mx)$ is the reduced mass of the DM-nucleon system, with nucleon mass \(m_{\rm N} \simeq 1\) GeV and nucleon form factor \(f_{\rm N} = 0.30\). Using Eq.~\ref{directdet}, we extract the cut-off scale \(\Lambda\) from experimental bounds and project these constraints onto the \((\Lambda^{-1}, \mx)\) plane. The shaded regions in Fig.~\ref{detection} correspond to the exclusion limits from different experiments. The green and purple shaded areas represent the current constraints from XENONnT and LZ, respectively, which probe DM masses up to $\mathcal{O}(3)$ TeV. The next-generation experiment XLZD (orange shaded) aims to further improve sensitivity, reaching the so-called "neutrino fog" limit \cite{Billard:2021uyg}. Additionally, for $\mx<\mh/2$ GeV, the invisible Higgs decay measurements at the LHC provide the constraints on $\lmd$, shown by the magenta-shaded regions. We take the invisible branching ratio ($\rm{BR}_{\rm inv}=\Gamma_{h\rightarrow\rm \chi\chi}/(\Gamma_h+\Gamma_{h\rightarrow\rm\chi\chi})$) to be $\rm{BR}_{\rm inv} \lesssim 0.11$ at $95\%\,$CL \cite{ATLAS:2022yvh,CMS:2023sdw,ATLAS:2023tkt}. However, for comparison, we show the viable DM parameter space for three choices of low reheating temperature $\tre=(\tbbn,0.5,10)$ GeV and for the higher $\tre$, required larger couplings (see Fig.~\ref{fullpara}). Notably, all these curves lie well below the sensitivity reach of current and near-future direct detection experiments, implying that the DM candidates considered in our scenario remain undetectable with existing technology. To probe this parameter space experimentally, significant improvements in detection sensitivity would be required.
\section{Conclusion}\label{conclusion}
In this work, we discussed the impact of reheating on the production of Higgs portal Majorana fermionic DM for both freeze-in and freeze-out scenarios. We examined how the background evolution of the inflaton field and its decay products influence DM production, considering the effects of EWSB. We assumed that the Universe is reheated via the perturbative decay of the inflaton into SM fermions. The thermal evolution during reheating strongly depends on the inflaton EoS (\(\wphi\)), which in turn depends on the shape of the inflaton potential. Therefore, DM production depends not only on the temperature evolution during reheating but also on the shape of the inflaton potential. In our analysis, we found that EWSB plays a crucial role in DM freeze-in and freeze-out. Specifically, we investigated whether it is possible to identify the parameter space region where the saturation of the DM yield took place before or after EWSB. The saturation of DM production before or after EWSB depends not only on the DM mass but also on the inflaton EoS , the thermal evolution of the Universe, and DM-SM interactions. We found that if \(\wphi > 3/7\), EWSB has no impact on freeze-in DM production. However, for \(\wphi < 3/7\), if the reheating temperature \(\tre < \tew\), freeze-in occurs after EWSB, requiring the DM mass to be below \(\mathcal{O}(500)\) GeV. Similarly, if freeze-out occurs after EWSB, the WIMP mass must be below \(\mathcal{O}(4)\) TeV for any \(\wphi\) and \(\tre\).\\
We also analyzed the relic-satisfied DM parameter space, characterized by the DM mass \(\mx\) and cutoff scale \(\Lambda\). In Fig.~\ref{fullpara}, we present the complete DM parameter space along with the relevant constraints. Finally, in Fig.~\ref{detection}, we compare our DM parameter space with the sensitivity of both current and future DM detection experiments. We find that the allowed parameter space lies well below the reach of existing and near-future direct detection experiments. Exploring this unexplored region would require substantial improvements in detection sensitivity.


Finally, we note that our numerical analysis was performed with \(\alpha=1\), although our results and discussions remain insensitive to the parameter \(\alpha\) within the limits set by recent Planck+BICEP/Keck observations \cite{Ellis:2021kad,Chakraborty:2023ocr}. Although these observations allow a broad range of $\alpha$, the reheating dynamics and DM phenomenology are independent of the choice of $\alpha$ within this allowed range. Different choices of $\alpha$ primarily influence inflationary observables such as the spectral index ($n_{\rm s}$) and tensor-to-scalar ratio ($r$). Using the latest constraints on the inflationary observables ($n
_{\rm s},r$) from Planck+BICEP/Keck \cite{Tristram:2021tvh,BICEP:2021xfz}, one can derive bounds on \(\tre\) \cite{Chakraborty:2023ocr}. For instance, in the case of \(\wphi=0\) and \(\alpha=1\), values of \(\tre\) below \(\mathcal{O}(10)\) GeV are disfavored at the $95\%$ confidence level (C.L) \cite{Ellis:2021kad,Chakraborty:2023ocr}. Consequently, the DM parameter space shown in Fig.~\ref{fullpara} for \(\tre\lesssim10\) GeV is not favorable from these observations. However, for certain values of \(\alpha\) within the allowed range, the entire range of \(\tre = (10^{15}-\tbbn)\) GeV remains permissible, allowing the full DM parameter space to be utilized. The same analysis also holds for other values of EoS $\wphi=1/3\,,1/2$.

\section{Acknowledgement}
RM would like to express his gratitude to Debaprasad Maity. RM also acknowledges the Ministry of Human Resource Development, Government of India (GoI), for financial support. SM extends sincere gratitude to the Department of Science and Technology (DST) for their generous financial support through the INSPIRE scheme (Fellowship No IF190758). TY expresses gratitude to Junpei Ikemoto.

\appendix
\section{Calculation the initial condition for reheating}\label{initcal}
For reference, we consider the well known $\alpha$-attractor $E$-model \cite{Kallosh:2013hoa,Ferrara:2014cca,Ueno:2016dim} inflaton potential,
 \bea
\label{pot1}
V(\phi)=\lambda^4\left(1-e^{-\sqrt{\frac{2}{3\alpha}}\frac{\phi}{M_p}}\right)^{2n}\,,
\eea
the potential scale $\lambda$ is uniquely determined by the CMB spectral index $\ns$, and scalar power-spectrum $A_{\rm s}$. We can write the scalar spectral index $\nss$ and the tensor-to-scalar ratio $r$ in terms  slow-roll parameters for a particular pivot scale $k_\star$, 
\begin{equation}{\label{eq18}}
\begin{aligned}
&\nss=1-6\,\epsilon(\phi_\star)+2\,\eta(\phi_\star)=1-\frac{8\,n}{3\,\alpha}\frac{\left(n+e^{\sqrt{\frac{2}{3\alpha}}\frac{\phi_\star}{M_p}}\right)}{\left(e^{\sqrt{\frac{2}{3\alpha}}\frac{\phi_\star}{M_p}}-1\right)^2}\,,\\
&r=16\,\epsilon(\phi_\star)=\frac{64\,n^2}{3\,\alpha}\frac{1}{\left(e^{\sqrt{\frac{2}{3\alpha}}\frac{\phi_\star}{M_p}}-1\right)^2}\,,
\end{aligned}
\end{equation}
where, the slow roll parameters are $\epsilon(\phi)=({\mp^2}/{2})({V (\phi)^\prime}/{V(\phi)})^2$ and $\eta(\phi)=\mp^2\, ({V (\phi)^{\prime\prime}}/{V (\phi)})$. After solving the above equations, we can write the field value $\phi_k$ at the horizon crossing, and the quantity $r$ in terms of $\nss,\,\alpha,\,n$, 
\begin{equation}
\begin{aligned}
    &\phi_\star=\sqrt{\frac{3\alpha}{2}}M_p \ln{\left[1+\frac{4n+\sqrt{16n^2+24\alpha n(1-\nss)(1+n)}}{3\alpha(1-\nss)}\right]}\,,\\
    &r=\frac{192 \, \alpha \, n^2 (1 - \nss)^2}{\left( 4n + \sqrt{16n^2 + 24\alpha n (1 - \nss)(1 + n)} \right)^2}\,.
    \end{aligned}
\end{equation}
The inflationary e-folding number $(\nk)$ between the exit of the horizon of the scale $k_\star$ at $\phik$, and the end of inflation
at $\phiend$, can be calculated in the slow-roll approximation,
\begin{equation}{\label{hk}}
\begin{aligned}
&\nk=\int^{a_{\mathrm{end}}}_{a_\star} d(\ln a)=\frac{1}{M_p}\int^{\phiend}_{\phi_{k}}\frac{d\phi}{\sqrt{2\epsilon}}=\frac{3\alpha}{4} \left( e^{\sqrt{\frac{2}{3\alpha}} \cdot \frac{\phi_k}{\mp}} - e^{\sqrt{\frac{2}{3\alpha}} \cdot \frac{\phi_{\mathrm{end}}}{\mp}}-\sqrt{\frac{2}{3\alpha}} \,\frac{(\phi_k-\phi_{\mathrm{end}})}{\mp}\right)\,.
\end{aligned}
\end{equation}
Here $\phi_k$ and $\phiend$ represent the values of the inflaton field at the point of horizon crossing and at the end of the inflation, respectively. The value of $\phi_{\mathrm{end}}$ is determined by the condition $\epv(\phi_{\mathrm{end}}) = 1$, marking the end of inflation. Using this condition, the field value and the potential at the end of inflation can be expressed as follows,
\begin{equation}{\label{phiend}}
 \phiend=\frac{\sqrt{3\alpha}}{2n}M_p\ln{\left(\frac{2n+\sqrt{3\alpha}}{\sqrt{3\alpha}}\right)}\quad,\quad V(\phiend)=\Lambda^4\left(\frac{2n}{2n+\sqrt{3\alpha}}\right)^{2n}\, .
\end{equation}
To find the potential scale $\lambda$, we use the following relation \cite{Kallosh:2013hoa},
\begin{equation}
    V(\phik)=\frac{3\,\pi^2\,A_{\text s}\,\mp^4}{2}\,r=\Lambda^4\left(1-e^{-\sqrt{\frac{2}{3\alpha}}\frac{\phik}{\mp}}\right)^{2\,n}\,.
\end{equation}
Using the expression of $\phik$ and $r$, we obtain,
\begin{eqnarray}\label{eq:elambda}
 \lambda=\mp\,\left(\frac{3\pi^2rA_s}{2}\right)^{1/4}\left[\frac{2n(2n+1)+\sqrt{4n^2+6\alpha(1+n)(1-\nss)}}{4n(1+n)}\right]^{n/2} \,.
 \end{eqnarray}
Finally, using this expression of $\Lambda$ in Eq.~\ref{phiend}, we can determine the initial condition for reheating, 
\begin{equation}
    \rho_\phi^{\rm end}\simeq\frac{3}{2}\,V(\phiend)=\frac{9\,\pi^2\,r\,\As}{4}\,\mp^4\left(\frac{2n}{2n+\sqrt{3\alpha}}\right)^{2\,n}\left[\frac{2n(2n+1)+\sqrt{4n^2+6\alpha(1+n)(1-\nss)}}{4n(1+n)}\right]^{2\,n}.
\end{equation}
Throughout our work, we have used $k_\star=0.05\,\mbox{Mpc}^{-1}$  as pivot scale, $A_{\rm s}\simeq 2.1\times 10^{-9}$, and $\ns=0.9649$ for the CMB observations from Planck \cite{Planck:2018vyg}.
\section{Upper bounds on inflationary scale and reheating temperature from CMB}\label{upperh}
The relation between inflationary scale $H^{\rm CMB}_{\rm I}$ and the tensor-to-scalar ratio $r$ is \cite{Kallosh:2013hoa},  
\begin{equation}{\label{hk}}
\begin{aligned}
&H^{\rm CMB}_{\rm I}=\frac{\pi\,\mp\sqrt{r\,A_{\rm s}}}{\sqrt 2}\,,
\end{aligned}
\end{equation}
Applying the recent bound on $r\leq0.035$ from Planck \cite{Planck:2018vyg}, using this bound in Eq.~\ref{hk}, we obtain an upper bound on inflationary scale,
\begin{equation}
    H^{\rm CMB}_{\rm I}\leq2\times10^{-5}\,\mp
\end{equation}
During inflation, the Hubble parameter remains nearly constant. At the end of inflation, we can therefore take $\hend=H^{\rm CMB}_{\rm I}$.
The Hubble parameter at the end of reheating can be written as, 
\begin{equation}
   \hre\simeq\frac{\pi}{3}\sqrt{\frac{g_\star}{10}}\frac{\tre^2}{\mp}\,. 
\end{equation}
Using the fact, $\hre\leq\hend$, and it gives,
$\tre\leq5\times10^{15}$ GeV. This is the maximum possible upper bound of reheating temperature.

\section{Lyman-$\alpha$ bound for FIMP}\label{Lymanbound}
At the time of freeze-in, DM particles possess a large momentum, typically of the order of the thermal bath temperature. Due to this large initial momentum, DM particles can have a large free-streaming length, potentially suppressing structure formation on small scales. If DM has no self-interactions, its momentum simply redshifts with the expansion of the Universe after freeze-in. The present-day momentum $\po$ is related to the momentum at freeze-in $\pfi$ by
\begin{equation}
    \po=\frac{\afi}{\ao}\pfi\,.
\end{equation}
Since DM particles are produced from the thermal bath, their typical momentum at freeze-in is $\pfi\simeq3\,\tfi$. If the freeze-in occurs after reheating, then 
\begin{equation}
    \po\simeq\left(\frac{g_{\rm\star s}(\to)}{g_{\rm\star s}(\tfi)}\right)^{1/3}\left(\frac{\to}{\tfi}\right)\,\pfi=3\,\to\,\left(\frac{g_{\rm\star s}(\to)}{g_{\rm\star s}(\tfi)}\right)^{1/3}
\end{equation}
A lower bound on the DM mass can be derived from constraints on the present-day velocity of warm dark matter (WDM). Taking, $v_{\rm wdm}\lesssim1.8\times10^{-8}$ \cite{Masina:2020xhk} for $m_{\rm wdm}\gtrsim3.5$ keV \cite{Irsic:2017ixq}. The lower bound of the DM mass,
\begin{equation}
\mx\gtrsim1.67\times10^8\,\to\left(\frac{g_{\rm\star s}(\to)}{g_{\rm\star s}(\tfi)}\right)^{1/3}\,,
\end{equation}
where \( g_{\rm\star s} \) represents the number of relativistic degrees of freedom that contribute to the entropy of the SM. Taking \( g_{\rm\star s}(T_0) = 3.90 \) and considering \( g_{\rm\star s}(\tfi) \) within the range \( 106.75 \) (for high-temperature production) to \( 3.90 \) (for late time production), we find the corresponding Lyman-$\alpha$ bound $\mx\gtrsim(1.30-3.95)\times10^{-5}$ GeV. We have used this bound for \(\wphi<3/7\), since Lyman-\(\alpha\) provides the lower bound on the DM mass, and this mass freezes in after reheating if \(\wphi<3/7\).

On the other hand, if the freeze-in occurs during reheating, then 
\begin{equation}
    \begin{aligned}
        \po&=\frac{\afi}{\ath}\,\frac{\ath}{\ao}\,\pfi\\
        &\simeq\left(\frac{g_{\rm\star s}(\to)}{g_{\rm\star s}(\tre)}\right)^{1/3}\,\left(\frac{\to}{\tre}\right)\left(\frac{\tre}{\tfi}\right)^{\frac{8}{3(1+3\,\wphi)}}\,\pfi
    \end{aligned}
\end{equation}
For $\wphi>3/7$, freeze-in always occurs during reheating at $\tfi\sim\tmax$. For, $\wphi=0.5$ and taking $\tre=(10^{15}-0.004)$ GeV, the lower bound of $\mx\gtrsim(10^{-5}-10^{-6})$ GeV.
\newpage
\bibliographystyle{apsrev4-1}
\bibliography{Rajeshreference}
\end{document}